\documentclass[fleqn,usenatbib]{mnras}
\usepackage{longtable}
\usepackage{newtxtext,newtxmath}
\usepackage{amsmath}
\usepackage[T1]{fontenc}
\usepackage[dvipsnames]{xcolor}

\usepackage{graphicx}	

\newcommand{\Msdb}{\mbox{$\rm M_{\mathrm{sdB}}$}}
\newcommand{\Msun}{M$_{\odot}$}

\newcommand{\Lsun}{L$_{\odot}$}
\newcommand{\lppr}{\stackrel{<}{\scriptstyle \sim}}
\newcommand{\lappr}{\raisebox{-0.4ex}{$\lppr$}}
\newcommand{\gppr}{\stackrel{>}{\scriptstyle \sim}}
\newcommand{\gappr}{\raisebox{-0.4ex}{$\gppr$}}
\newcommand{\mesa}{{\sc mesa}}
\newcommand{\sse}{{\sc sse}}

\title[sdB stars with MESA]{The mass range of hot subdwarf B stars from MESA simulations}

\author[E. Arancibia-Rojas et al.]{
Eduardo Arancibia-Rojas$^{1}$\thanks{E-mail: eduardo.arancibiar@postgrado.uv.cl},
Monica Zorotovic$^{1}$\thanks{E-mail: monica.zorotovic@uv.cl},
Maja Vu\v{c}kovi\'{c}$^{1}$,
Alexey Bobrick$^{2}$,
Joris Vos$^{3}$,
\newauthor
Franco Piraino-Cerda$^{1,4}$\\
$^{1}$Instituto de F\'isica y Astronom\'ia, Universidad de Valpara\'iso, Av. Gran Breta\~na 1111, Valpara\'iso, Chile\\
$^{2}$Technion - Israel Institute of Technology, Physics Department, Haifa, Israel 32000\\
$^{3}$Astronomical Institute of the Czech Academy of Sciences, CZ-25165, Ondřejov, Czech Republic\\
$^{4}$Departamento de F\'isica, Universidad T\'ecnica Federico Santa Mar\'ia, Avenida Espa\~na 1680, 2390123, Valpara\'iso, Chile
}

\date{Accepted 2023 December 14. Received 2023 December 14; in original form 2023 April 26}

\pubyear{2023}

\begin{document}
\label{firstpage}
\pagerange{\pageref{firstpage}--\pageref{lastpage}}
\maketitle

\begin{abstract}
Hot subdwarf B (sdB) stars are helium core burning stars that have lost almost their entire hydrogen envelope due to binary interaction. Their assumed canonical mass of $\Msdb\sim0.47$\,\Msun\,has recently been debated given a broad range found both from observations as well as from the simulations. Here, we revise and refine the mass range for sdBs derived two decades ago with the Eggleton code, using the stellar evolution code \mesa, and discuss the effects of metallicity and the inclusion of core overshooting during the main sequence. We find an excellent agreement for low-mass progenitors, up to $\sim2.0$\,\Msun. For stars more massive than $\sim2.5$\,\Msun\,we obtain a wider range of sdB masses compared to the simulations from the literature. Our \mesa\,models for the lower metallicity predict, on average, slightly more massive sdBs. Finally, we show the results for the sdB lifetime as a function of sdB mass and discuss the effect this might have in the comparison between simulations and observational samples. This study paves the way for reproducing the observed Galactic mass distribution of sdB binaries.
\end{abstract}

\begin{keywords}
stars: subdwarfs - binaries: general - stars: evolution - stars: mass-loss - Galaxy: evolution
\end{keywords}
 

\section{Introduction}
\label{sec:intro}

Hot subdwarf B (sdB) stars are core helium-burning stars on the horizontal branch with a very thin hydrogen layer. Due to the interaction with a binary companion they have lost almost their entire envelope during their earlier evolution, but managed to ignite helium after the envelope was ejected \citep[see][for a comprehensive review]{heber09}.

Most studies of sdB stars assume a canonical mass of $\Msdb\sim0.47$\,\Msun, which corresponds to the core mass for a $\sim$solar-mass giant at the tip of the red giant branch (RGB), where the helium core flash occurs. However, the observational constraints of the masses reveal a rather broad mass range. For example, \citet{fontaine2012} obtained an empirical range of $\Msdb\sim0.35-0.63$\,\Msun\,from asteroseismology of 15 pulsating sdBs, and a somewhat broader range of $\Msdb\sim0.29-0.63$\,\Msun\,if eclipsing sdB binaries are also included. A similar mass distribution was recently found by \citet{Schaffenroth22} for a larger population of sdBs in close (post-common envelope) binary systems with low-mass main sequence (MS) or brown dwarf companions, as well as by \citet{leietal2023} for single-lined hot subdwarf stars in  \textit{The Large Sky Area Multi-Object Fiber Spectroscopic Telescope spectra survey} (LAMOST).

Theoretically, the range of possible masses for sdBs was calculated by \citet{han2002}, who showed that the resulting mass depends on the progenitor mass and other assumed parameters and physical processes, such as metallicity, core overshooting, etc. While the derived masses were restricted to a rather small range of $\Msdb\sim0.45-0.48$\,\Msun\,(i.e. close to the canonical value) for low-mass progenitors with initial masses $\lappr\,1.3$\,\Msun, progenitors with larger initial masses can lead to smaller sdB masses (as low as $\sim0.32$\,\Msun) but also to larger masses if the progenitor was more massive than $\sim3$\,\Msun.

Having accurate and reliable constraints of the sdB masses, both from an observational and theoretical point of view, is crucial to test the evolutionary paths toward these stars. The work from \citet{han2002} was based on the stellar evolution code from \citet{Eggleton71}. We here redo these calculations and refine the grids using the most updated and flexible open source code \textit{Modules for Experiments in Stellar Astrophysics} \citep[\mesa\footnote{version -r15140},][]{paxton2011,paxton2013,paxton2015,paxton2018,paxton2019,jermyn23} to provide updated mass ranges for sdB stars, depending on the mass and metallicity of the progenitor, and the sdB phase duration. This code stands out due to its collaborative, open-source nature, which ensures it is continuously being tested and updated. It incorporates modern numerical techniques for efficient and accurate stellar evolution simulations, and its modular design and flexible microphysics enable customization and versatility, supporting a wide range of astrophysical processes.
The inlist files for \mesa\, are also included, allowing for the reproduction of our calculations across a wide range of initial masses, metallicities, and overshooting strengths. This is particularly valuable for further binary population synthesis models and comparison with observational distributions.
We compare the results from \mesa\,with those obtained by \citet{han2002} with the Eggleton code, and study the effects of including core overshooting during the MS phase of the progenitors of the sdBs. 


\section{Models with MESA} 
\label{sec:model}

Previous works have already used the stellar evolution code \mesa\,to simulate sdBs. For example, \citet{schindler15} calculated a series of sdB stellar evolution models with \mesa\,to compare with observational results derived from spectroscopy \citep{green2008} and with previous sdB models that use different stellar codes \citep{Charpinet2002,Bloemen2014}. They assumed initial masses between 1.0 and 2.5\,\Msun\,and applied the built-in tool from \mesa\,called `Relax Mass' to artificially remove the envelope at the tip of the RGB, in order to reproduce the sdB phase. Later,  
\citet{Ghasemi2017} used \mesa\,models to study the effects of overshooting by comparing their models with the evolutionary and asteroseismic properties of the observed sdB pulsator KIC\,10553698. They modeled the evolution of a star with an initial mass of 1.5\,\Msun\,from the pre-main-sequence phase until the core helium depletion, and also used the `Relax Mass' tool at the tip of the RGB to remove the envelope in order to obtain an sdB star, with different values of the overshooting parameter.

Following these works, we here used the single star module from \mesa\,to evolve stars with different initial masses and metallicities, and 
the `Relax Mass' tool to artificially remove the envelope during the RGB. 
The maximum sdB mass for a given progenitor was obtained by removing the envelope at the tip of the RGB, defined by the onset of helium burning in the core. The minimum sdB mass, on the other hand, was obtained by removing the envelope at different stages after the MS, but before the tip of the RGB, in order to find the minimum mass that manages to ignite helium in the core after removing the envelope. For low-mass progenitors, in which the core becomes degenerate during the RGB phase and ignition of helium in the core occurs explosively (experiencing core helium flash), we further required that stable helium burning was achieved, thus leading to a developed carbon core. Otherwise, the star was considered as a `failed sdB star'. 

We used a grid of initial masses from $0.8$ to $6$\,\Msun\,and two different metallicities \citep[$Z = 0.02$ and $0.004$, following][]{han2002}. Core overshooting was included during the MS for both metallicities (see Sec.\,\ref{sec: PreMS_to_TAMS}). Models without overshooting where also considered but only for $Z = 0.02$, in order to allow for a direct comparison with the same models from \citet{han2002}. 
For each possible progenitor in our grid, the modeling process was divided into three steps: i) evolving the star from the pre-main-sequence phase to the terminal age main sequence (TAMS); ii) evolving the star from the TAMS to the tip of the RGB; iii) loading the star at different evolutionary stages after the TAMS and rapidly removing the envelope. In what follows, we detail some important parameters considered in each of these steps.

\subsection{From pre-MS to TAMS}
\label{sec: PreMS_to_TAMS}
The predictive mixing scheme was used to establish the limit between the radiative and convective zones during the MS \citep[see Section\,2.1 in][]{paxton2018}.
The mixing length alpha ($\alpha_{MLT}$) parameter, defined as the local pressure scale height, was set to $1.8$ \citep[e.g.,][]{Ostrowski2021}. 

The inclusion or not of core overshooting is important during the MS, because the mass of the helium core at the TAMS depends on mixing processes, therefore affecting the derived mass range for the sdBs. There is growing evidence that models without overshooting do not reproduce observations for giant stars initially more massive than $\sim2$\,\Msun\,\citep[e.g.,][]{constantino+baraffe2018}. Therefore, we decided to include overshooting in our standard \mesa\,models. Some models without overshooting were also calculated for $Z = 0.02$, in order to analyse the effect that overshooting has on the sdB mass, but also to allow for a more direct comparison with the masses obtained with the Eggleton code, presented by \citet{han2002}. 

In our standard models, where core overshooting was included during the MS phase, we adopted the exponential diffusive overshooting scheme from \mesa\,\citep{herwig2000,paxton2011}, in which overshooting is treated as a diffusion process with an exponentially decreasing diffusion coefficient \citep[see, e.g.,][]{zhang2022}. The free parameter $f_{\rm ov}$ sets the extent of the overshoot region. Here we adopted the recommended value from \citet{herwig2000},  $f_{\rm ov}=0.016$, based on fits to the stellar models from \citet{schaller92}, which is roughly equivalent to $f_{\rm ov,step} = 0.2$ in the step overshoot scheme (also implemented in \mesa).

The models were saved when the TAMS was reached, defined by the depletion of hydrogen in the center (when the central lower limit for hydrogen is $10^{-4}$).

\subsection{From TAMS to the tip of the RGB}

The total power of the helium burning reactions was used to stop the simulation at the tip of the RGB, with the limiting value set to $10$\,\Lsun\footnote{Based on MESA test suite 1M\_pre\_ms\_to\_wd.}.
This condition corresponds to the rapid increase in helium burning luminosity following helium ignition at the tip of the RGB.
For the minimum sdB mass the models should be stopped before reaching the tip of the RGB. This was obtained simply by stopping the simulation at a specific model number, which corresponds to a helium core mass smaller than the core mass at the tip of the RGB. Whether this core mass ignites helium after removing the envelope was tested in the third step of the modelling process.

\subsection{Removing the envelope after the TAMS and evolving until the white dwarf cooling track}

For each progenitor, we loaded a previously saved model either from the tip of the RGB (to derive the maximum sdB mass), or with different helium core masses after the TAMS (to derive the minimum sdB mass) and applied the `Relax Mass' process from \mesa\,to remove the envelope. A small hydrogen envelope was left around the helium core, part of which was incorporated into the core before helium ignition. We found that leaving an envelope of $0.01$\,\Msun\,for the `Relax Mass' process resulted in a hydrogen envelope mass of the order of $\sim10^{-2}-5\times10^{-4}$\,\Msun\,during the sdB phase \citep[e.g.,][]{schindler15}. 

From this point, having already removed the envelope, we allowed the star to evolve until the luminosity drops below $\log \left( L/L_\odot \right) = -3.5$, when the star is already on the white dwarf cooling track. 
We clarify here that, while we were not interested in the white dwarf phase, setting such a low luminosity as the stopping condition in \mesa\, was needed to ensure that we really obtain the minimum sdB masses. This is because, as we will see in the next Section, when the envelope is removed before the tip of the RGB, the star first needs to contract, following the white dwarf cooling track, until enough compression allows for helium ignition.

\vspace{0.5cm}
\noindent An example of a \mesa\,inlist file for each of the three steps in the modelling process can be found in Appendix\,\ref{A:inlists}.

\subsection{Finding the minimum sdB masses}
\label{subsec:findmin}

To evaluate whether a model has succeeded in burning helium in a stable way, i.e. going through the sdB phase, we looked at the mass of the carbon core. If the model had a final carbon core mass larger than $0$ it was assumed that it experienced a phase of core helium burning as an sdB star. 
We chose this condition, instead of a condition based on helium burning, because some models that managed to ignite helium under degenerate conditions did not reach a stable helium burning phase and moved directly to the white dwarf cooling sequence. In those `failed sdBs', ignition most likely did not reach the center and the remnant was a helium-core white dwarf, i.e. with a final carbon-core mass of $0$\footnote{We clarify that the carbon-core mass in \mesa\,can be $0$ while the carbon concentration is not. This is because the radius of the carbon core in \mesa\,is defined as the outermost location where helium-4 mass fraction is $\leq0.01$ and carbon-12+oxygen-16 mass fraction is $\geq0.1$. Therefore, if helium concentration in the center is larger than 1\%, the effective carbon core mass is zero.}.

Initially, we applied the `Relax Mass' process loading models after the TAMS with increasing core masses, in steps of $0.01$\,\Msun, until we found a model that went through the sdB phase. Later, we used finer steps of $0.001$\,\Msun\,between this core mass and the previous model, that did not create an sdB star, in order to get an accurate minimum sdB mass.


\section{Results} 
\label{sec:res}

Here we present the results we obtained with the \mesa\,code. First, we focus on the details of the evolution for a given initial mass all the way to the white dwarf cooling track for the two limiting cases, i.e. for the maximum and minimum sdB mass. Then we present and discuss the sdB mass ranges obtained in this work. 

\subsection{Evolution before and during the helium ignition}

\begin{figure*}
\begin{center}
  \includegraphics[width=0.75\textwidth]{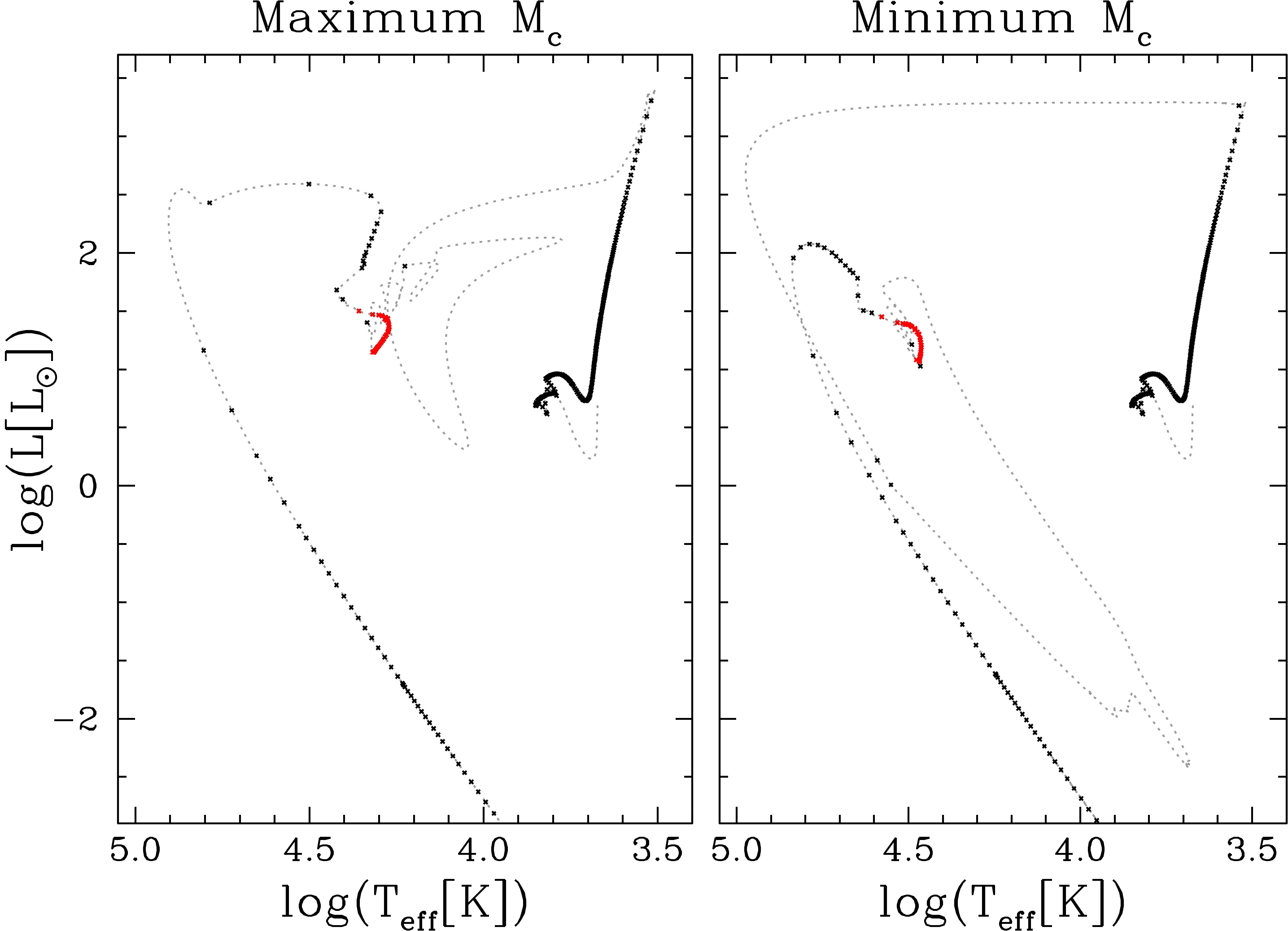}
  \caption{HR diagram evolution for a $1.5$\,\Msun\,star where the envelope was artificially removed either at the tip of the RGB (\textit{left}) or when the core mass on the RGB corresponds to the minimum mass that ignites helium after the envelope is ejected (\textit{right}). The gray dotted lines show the whole evolution while the black and red dots are in steps of $1$\,Myr with the sdB phase highlighted as red dots.}
    \label{FIG:mesaHR}
\end{center}
\end{figure*}

Figure\,\ref{FIG:mesaHR} shows an example of the evolution in the HR diagram for a typical sdB progenitor with an initial mass of $1.5$\,\Msun, $Z = 0.02$ and including core-overshooting during the MS phase. The left panel shows the evolution when the envelope was artificially removed at the tip of the RGB phase, which results in the maximum sdB mass for this initial mass ($\Msdb=0.4708$\,\Msun). In the right panel, on the other hand, the envelope was removed when the core mass was the minimum needed to ignite helium after losing the envelope, leading to the minimum sdB mass for the same progenitor mass ($\Msdb=0.449$\,\Msun). The gray dotted lines show the whole evolution, while the black and red dots correspond to steps of $1$\,Myr and the red dots are highlighting the sdB phase.

We can see that by removing the envelope at the tip of the RGB the star manages to ignite the helium core very quickly, because the necessary conditions for mass, pressure and temperature were already almost reached. On the other hand, for the minimum mass, the star first needs to contract and heat to reach the necessary conditions to ignite helium. This can be seen in the right panel of Fig.\,\ref{FIG:mesaHR}, where the star first goes towards the white dwarf cooling track until enough compression is achieved and helium can be ignited, moving the star back up in the HR diagram to become an sdB. A similar behaviour was obtained by \citet{byrne2018}.

\begin{figure*}
\begin{center}
  \includegraphics[width=0.75\textwidth]{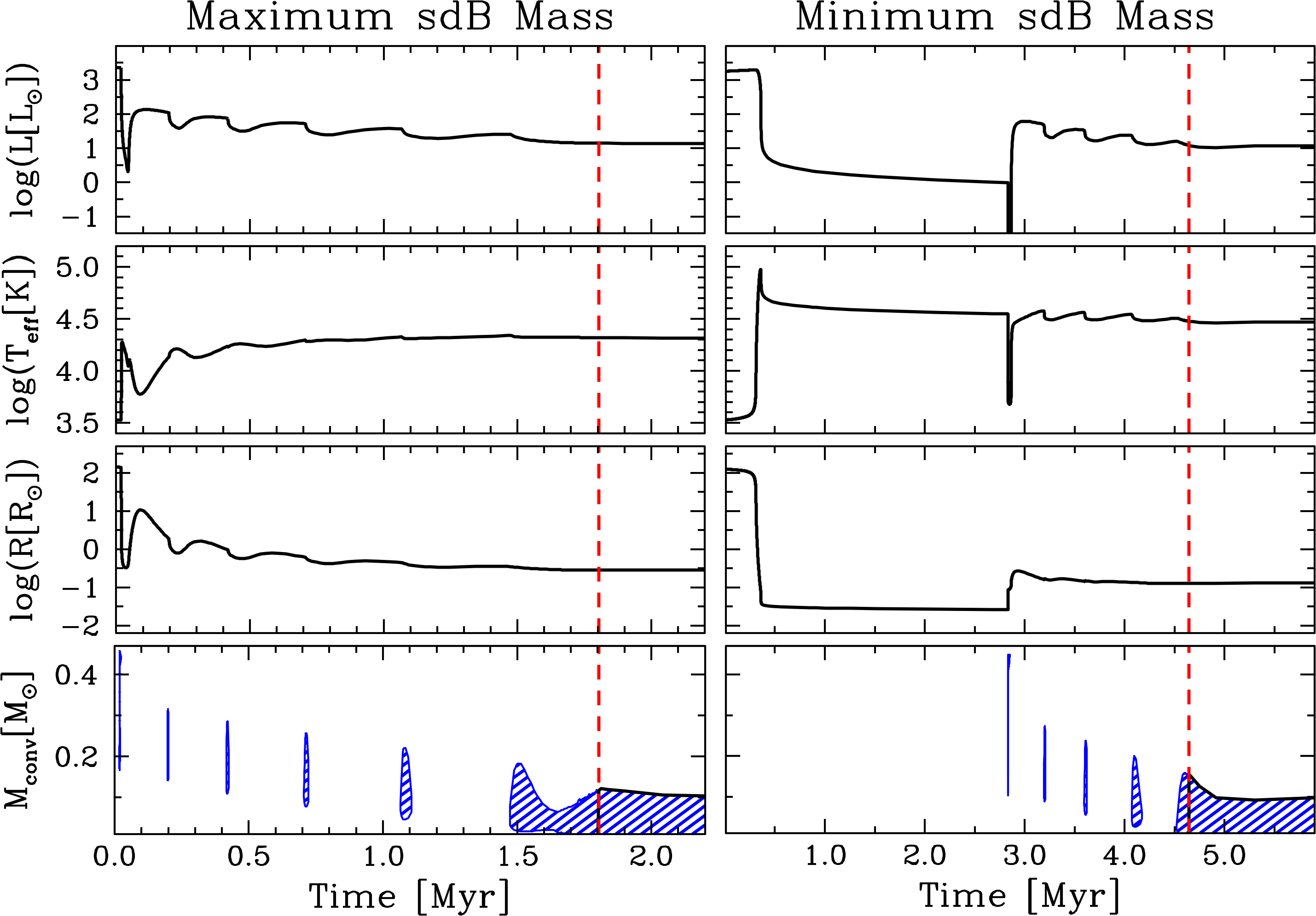}
  \caption{Evolution of the total luminosity (\textit{top)}, effective temperature (\textit{second panel from top}), radius (\textit{third panel from top}) and location of the largest convective region (\textit{bottom}) for a $1.5$\,\Msun\,star after removing the envelope at the tip of the RGB (\textit{left panel}) or when the core mass on the RGB corresponds to the minimum mass that ignites helium after the envelope is ejected (\textit{right panel}). In the bottom panel, the black line represents the mass of the convective core while the blue dashed regions show the location of the largest convective zone. The time was set to $0$ after the envelope was fully removed to facilitate visualization.
  We show only the first few Myr after the envelope ejection, to focus on the phase of helium flashes, where the convection region is approaching the center with each flash. The red dashed line indicates the time when convection reaches the center, setting the beginning of the sdB phase.}
    \label{FIG:flashes}
\end{center}
\end{figure*}

\begin{figure*}
\begin{center}
  \includegraphics[width=0.75\textwidth]{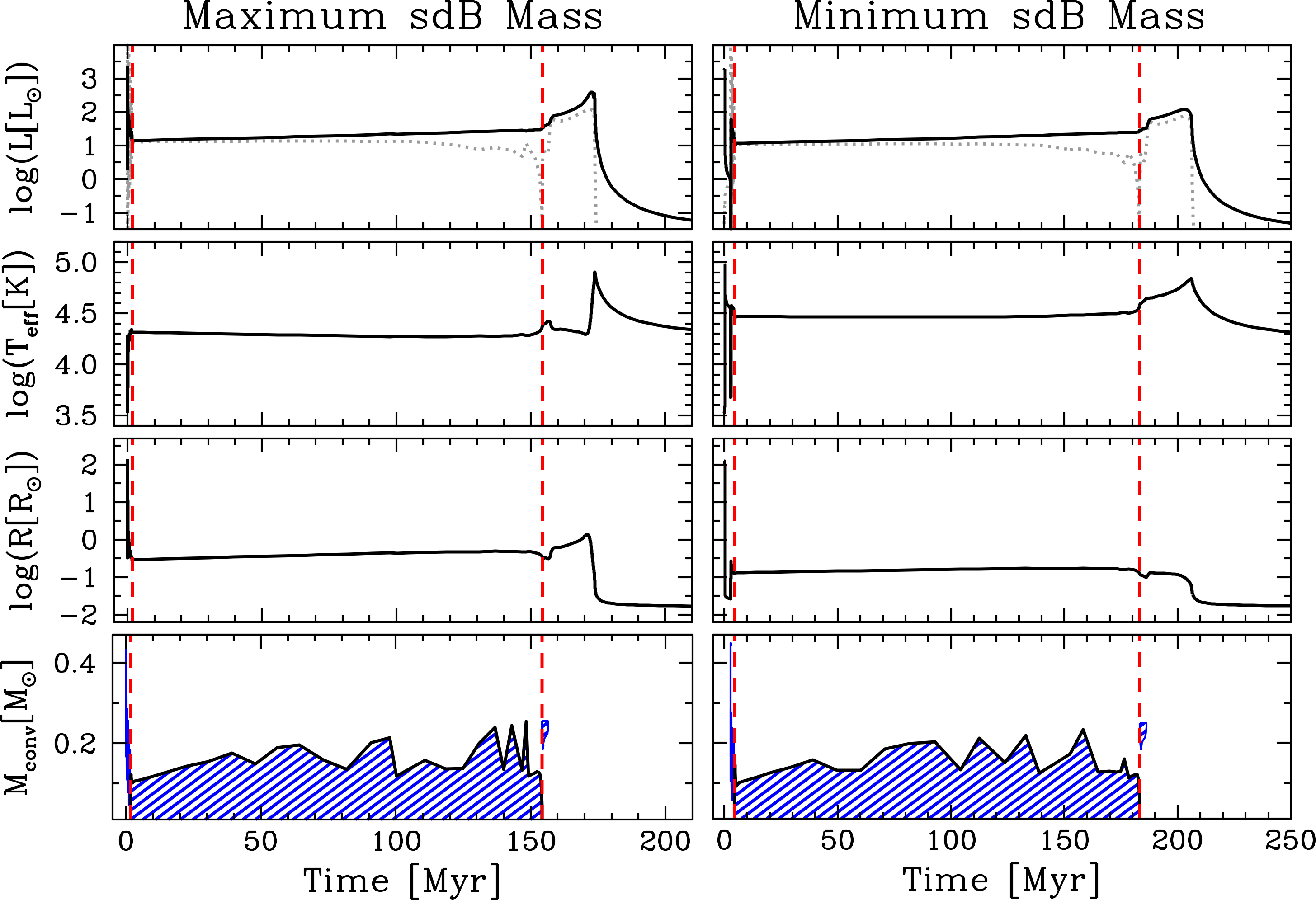}
  \caption{Same as in Fig.\,\ref{FIG:flashes} but including the sdB and post-sdB phases. The helium luminosity is also included in the top panel as a dotted gray line. The two dashed red lines indicate the beginning and end of the sdB phase.}  
    \label{FIG:history3}
\end{center}
\end{figure*}

The sdB phase corresponds to the phase of stable helium burning in the core of the naked star, which was identified due to the presence of a convective core after the removal of the envelope. Before helium ignition, the core is either degenerate or radiative (for low- and high-mass progenitors, respectively). The ignition of helium in the core provides a sufficient energy flow to ensure convection. When helium at the very center is depleted the core again becomes radiative (and later degenerates).

Given the mass of the initial star in these models, helium is ignited under degenerate conditions in a series of rapid helium flashes that cause the loops in the luminosity and effective temperature seen in Fig.\,\ref{FIG:mesaHR} just before the sdB phase. 
In Fig.\,\ref{FIG:flashes} we show the evolution of the luminosity (top), effective temperature (second panel from top), radius (third panel from top) and location of the largest (by mass) convective region (bottom) for the $1.5$\,\Msun\,star after the envelope was removed either at the tip of the RGB (left panel) or when the core mass corresponds to the minimum mass that ignites helium after the envelope is ejected (right panel), focusing on the first few Myr after removing the envelope, to see the behaviour of important stellar parameters during the helium flashes. The blue dashed regions in the bottom panel show the location of the largest convective zones while the black line represents the convective core mass. From the bottom panel of this figure, it can be observed that the ignition of helium starts off center, generating a series of flashes and the development of a convective zone that moves towards the center of the star with each subsequent flash. Each of these flashes causes a contraction of the whole star and a drop in luminosity, while the effective temperature increases. The only exception is for the first flash in the model with the minimum sdB mass (right panel), because the star was already a white dwarf when the first flash occurred, and helium ignition resulted in an increase in radius and luminosity, and a decrease in effective temperature. 
The effect of the flashes on the surface are more evident for the first and more external flash, while the subsequent ones are progressively less intense, closer to the center, of longer duration and with less effect on the surface of the star. For this particular case, ignition of helium reaches the center during the sixth flash for the sdB with the maximum mass (left panel) and during the fifth flash for the sdB with the minimum mass (right panel), setting the beginning of the sdB phase (vertical red dotted line) when a convective core appears. These results are strongly consistent with those obtained by \citet[][their Fig.\,C1]{Ostrowski2021} for a $1$\,\Msun\,star after removing the envelope at the tip of the RGB.  

In Fig.\,\ref{FIG:history3}, we show the same four panels as in Fig.\,\ref{FIG:flashes} but for the whole sdB and post-sdB phases until the star is again on the white dwarf cooling track. We also show in the top panels, the power of helium burning (as the logarithm of the total thermal power from triple-alpha process, excluding neutrinos, in solar luminosities), given by the dotted gray lines. 
The two red dashed vertical lines indicate the beginning and end of the sdB phase, which last $\sim150$\,Myr and $\sim180$\,Myr for the models with the maximum and minimum sdB mass for the given progenitor mass, respectively. 
Both the mass of the convective core as well as the duration of the sdB phase are also consistent with the results obtained by \citet[][their Fig.\,6]{Ostrowski2021}. They found that for an initial mass of $1.0$\,\Msun, and using the predicting mixing scheme, the duration of the sdB phase was $147.9$\,Myr when the envelope was removed at the tip of the RGB.

We can see from Fig.\,\ref{FIG:history3} that the sdB that descends from the tip of the RGB phase (left panel) has a larger radius and a lower effective temperature, compared to the sdB with the minimum mass (right panel) for a $1.5$\,\Msun\,progenitor. This is a consequence of the hydrogen envelope that remains during the sdB phase. Although in both cases we left the same hydrogen envelope of $0.01$\,\Msun\,around the helium-core during the `Relax Mass' process, the sdB that descends from the tip of the RGB ignites helium very quickly (only $\sim17\,000$\,yr after the envelope removal), while hydrogen was still being burned and incorporated into the core. The ignition of helium halts the burning of hydrogen and results in an sdB with a larger hydrogen envelope (of $\sim9\times10^{-3}$\,\Msun). On the other hand, for the model with the minimum sdB mass, the hydrogen burning phase lasts for $\sim140\,000$\,yr after the envelope removal, while ignition of helium occurs $\sim2.8$\,Myr after removing the envelope, when the star was already a white dwarf and the hydrogen envelope was smaller ($\sim5\times10^{-4}$\,\Msun\,for this model). 

It can also be seen from this figure that after the end of the sdB phase, when helium core burning stops, the whole star experiences a rapid contraction phase, decreasing the radius while increasing the luminosity and effective temperature. The power of helium burning quickly recovers (see dotted gray line in the top panels), increasing the luminosity, due to the ignition of helium in a shell around the inert core (small convective zone just after the end of the sdB phase in the bottom panels). This shell is initially convective but quickly becomes radiative, as the radius increases.
The behaviour of the radius and effective temperature during the shell helium burning phase is very different for the two models. For the maximum sdB mass, the radius increases causing the effective temperature to decrease slightly. The star moves up and slightly to the right on the HR diagram while remaining in the same temperature range of B-type stars, due to the thicker hydrogen envelope.
On the other hand, for the minimum sdB mass, the radius decreases slightly and the effective temperature increases, moving the star up and to the left on the HR diagram, ascending towards the phase where hot subdwarf O (sdO) stars reside. 
During the helium shell burning phase, part of the remaining hydrogen envelope is burnt and converted to helium, leading to very similar hydrogen envelope masses ($\sim4\times10^{-4}$\,\Msun) for the resulting white dwarfs in both cases. However, given that there was a larger hydrogen envelope around the more massive sdB, a peak in the luminosity and a rapid increase in effective temperature, associated with hydrogen shell burning, is observed in the left panel just before entering the white dwarf cooling track. 
Given the small envelope left, the stars fail to reach the asymptotic giant branch (which is true for all the sdBs we simulated, regardless of the progenitor mass). The post-sdB phase, with shell helium burning, lasts for $\sim10-20$\,Myr, after which the power of helium burning drops dramatically, considerably decreasing the luminosity, effective temperature and radius, following the cooling track of a carbon/oxygen or a hybrid helium/carbon/oxygen white dwarf \citep[see, e.g., ][for a discussion on the formation of hybrid white dwarfs]{zenati19}. The formation of hybrid white dwarfs with helium shells larger than $0.01$\,\Msun\,are of crucial interest for modelling the transient resulting from merging white dwarf binaries \citep{Perets19}. 

We note that the differences just outlined for the two sdBs, with the maximum and minimum sdB mass for a $1.5$\,\Msun\, progenitor, are a direct consequence of having left the same envelope mass after the envelope removal, which might not be realistic. It might well be that removing the envelope closer to the tip of the RGB is more efficient, given the lower binding energy of a more extended envelope. This might translate into an initially smaller hydrogen envelope around the core compared to the case in which the envelope is ejected earlier on the star's evolution. Therefore, one should not conclude from our results that sdBs descending from more evolved progenitors are colder and larger due to their larger hydrogen envelope, or that they do not pass thought the location of sdO stars during the shell helium burning phase. Given that the detail mechanism of the mass loss to form an sdB is not entirely understood, the mass of hydrogen that remains around the helium core after removing the envelope is unknown, albeit it should be small. As we mention earlier, we have chosen to leave $0.01$\,\Msun\,following the work of \citet{schindler15}. 

It is out of the scope of this paper to review in detail the evolution for different progenitor masses, and the example just shown was meant to illustrate the results that can be obtained with \mesa\,for a typical progenitor mass. However, for more massive progenitors ($\gappr\,2.0-2.5$\Msun) helium ignites smoothly and at the very center of the star. Therefore, a convective core appears immediately after helium ignition. The rest of the evolution is similar to the cases we have shown here. However, we note that as the mass of the progenitor is increased, the difference between the minimum and maximum sdB mass is greater (as will be seen in Sec.\,\ref{sec:sdBmasses}). This leads to larger differences in the HR diagram during the sdB phase, especially with respect to the luminosity, which can differ by more than an order of magnitude (more massive sdBs being more luminous), while the effective temperature remains in a similar range. Furthermore, the duration of the sdB phase strongly depends on the mass of the sdB star, as we will discuss in Sec.\,\ref{sec:lifetime}.

\subsection{SdB masses}
\label{sec:sdBmasses}

\begin{figure}
\begin{center}
  \includegraphics[width=0.49\textwidth]{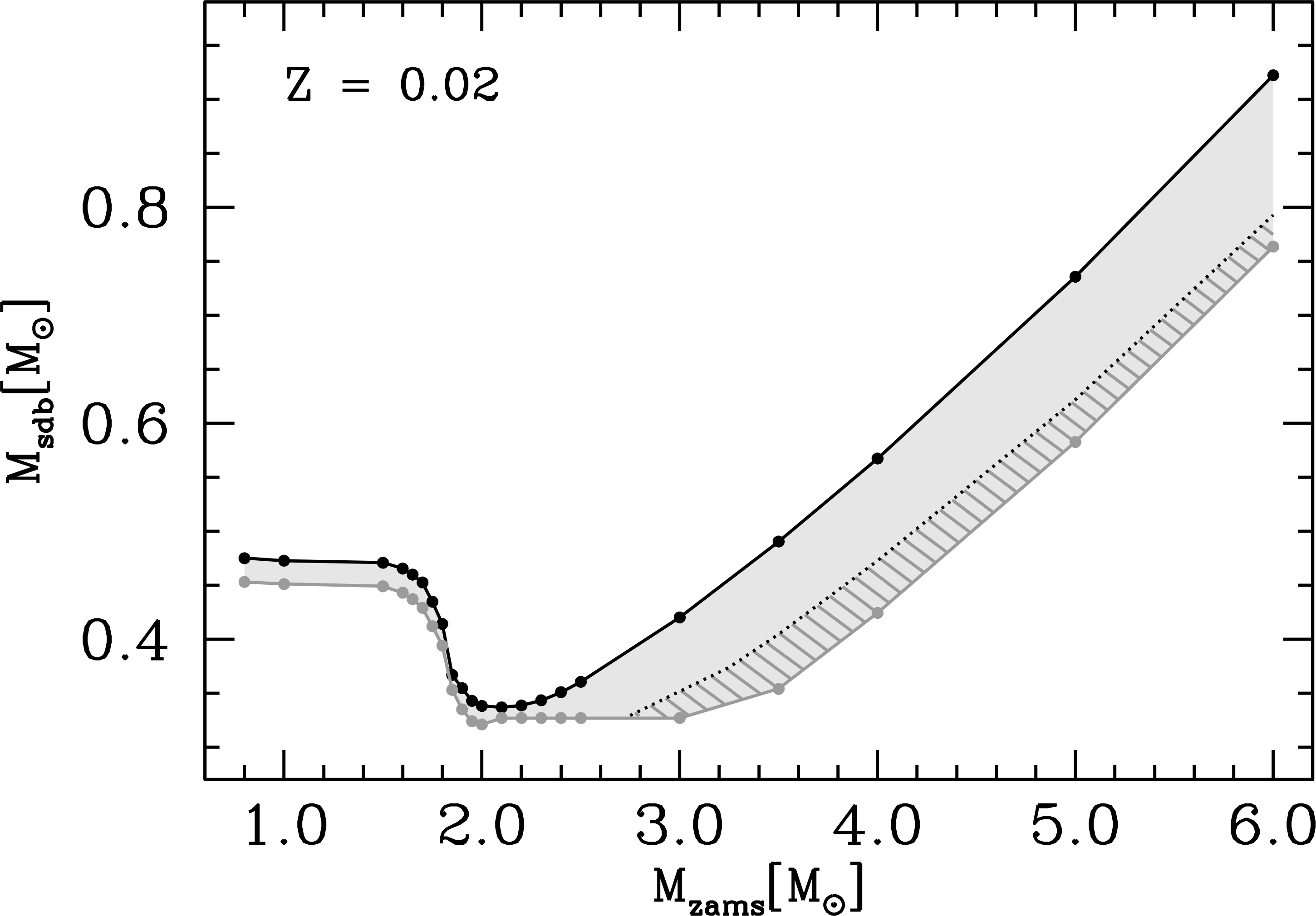}
  \caption{Maximum (black) and minimum (gray) sdB masses as a function of initial mass for the \mesa\,models with $Z = 0.02$ and core overshooting during the MS.  The gray area corresponds to the whole range of possible sdB masses, while the dashed area highlights sdBs descending from progenitors on the subgiant branch phase.}
    \label{FIG:range_002ov}
\end{center}
\end{figure}

Figure\,\ref{FIG:range_002ov} shows the maximum (black) and minimum (gray) sdB masses as a function of the zero age main sequence (ZAMS) mass for the models with $Z = 0.02$, with overshooting. The hydrogen envelope of $0.01$\,\Msun\,left around the helium-core during the `Relax Mass' process is considered as part of the sdB mass. As we explained in the previous section, a fraction of this hydrogen is burnt after the `Relax Mass' process, resulting in hydrogen envelopes of $10^{-2}-5\times10^{-4}$\,\Msun\,during the sdB phase. Typically, we obtained that a more massive hydrogen envelope remains around the sdB if the progenitor was closer to the tip of the RGB, i.e. for those that become sdBs faster after the envelope was removed. As we mentioned before, this is a direct consequence of leaving the same hydrogen envelope mass in the `Relax Mass' process for progenitors at different evolutionary stages, which might not be realistic. However, given that $0.01$\,\Msun\, is considered as the upper limit for the H envelope mass in sdBs \citep[e.g.,][]{heber16}, the sdB mass range will not be strongly affected by this.   

We note that the more massive stripped stars predicted here (with $\Msdb\,\gappr\,0.6$\,\Msun) will most likely be located in the region of the HR diagram that corresponds to sdOB or sdO stars, instead of sdBs \citep[see e.g., Fig.\,6 in][]{gotberg18}.

For stars with $\rm M_{\rm ZAMS} \sim 0.8 - 1.5$\,\Msun\,the maximum sdB mass is very close to the canonical value of $\sim0.47$\,\Msun. These stars have a high level of degeneracy in their cores during the RGB phase, which translates into a similar core mass at the tip of the RGB, regardless of the total mass. For larger progenitor masses, the maximum sdB mass decreases rapidly reaching a minimum value of $0.34$\,\Msun\, for $\rm M_{\rm ZAMS} \sim 2.0-2.1$\,\Msun. 
This is because the level of degeneracy in the core during the RGB phase decreases abruptly for ZAMS masses around $1.7-1.9$\,\Msun, which facilitates compression and heating of the core during the RGB, therefore allowing the star to reach the conditions for helium ignition at a smaller core mass. For initial masses above $\sim2.1$\,\Msun\, the level of degeneracy in the core can be neglected and the maximum sdB mass starts to increase for an increasing ZAMS mass as a direct consequence of the more massive core mass at the end of the MS.

\begin{figure}
\begin{center}
  \includegraphics[width=0.49\textwidth]{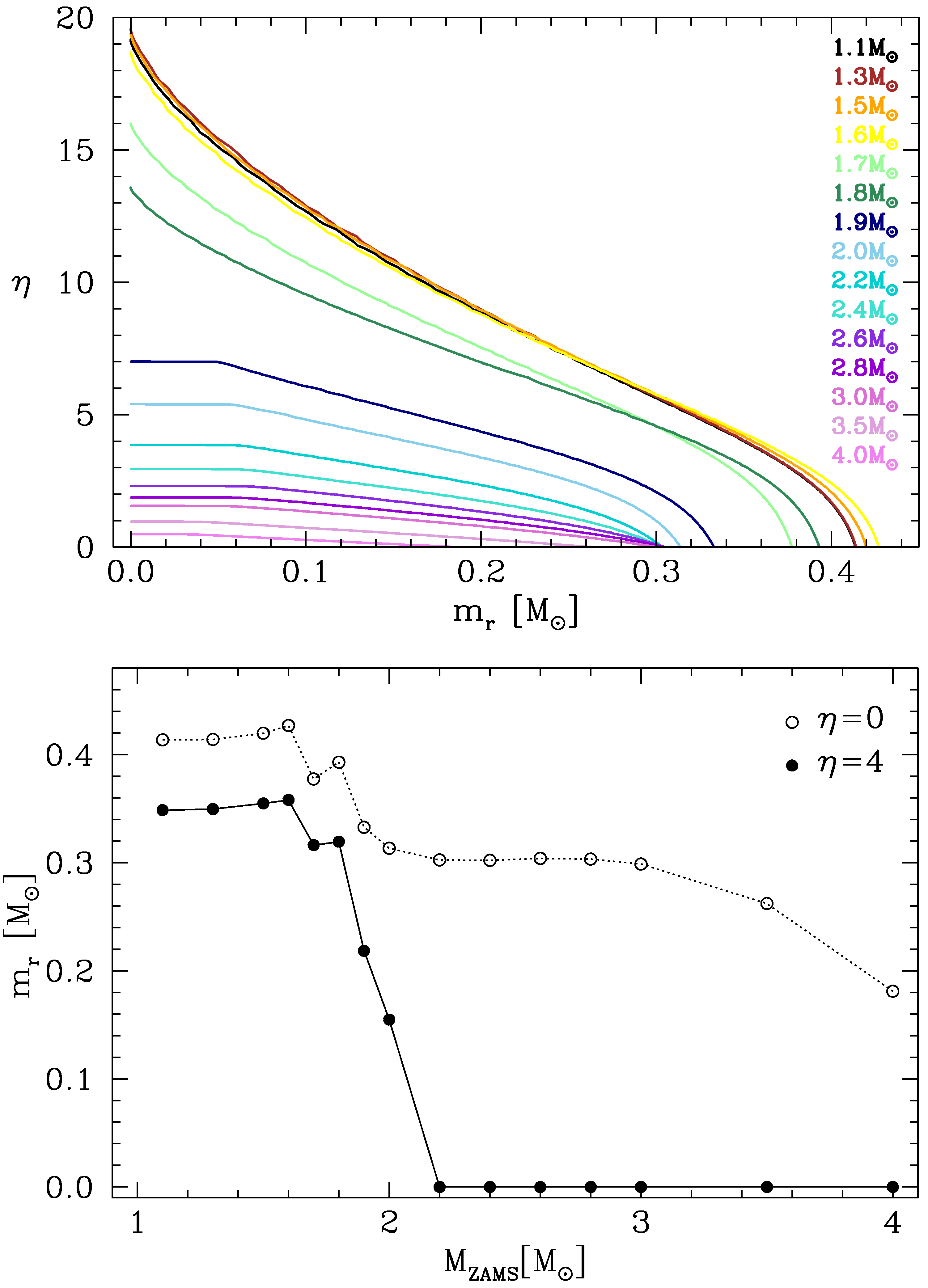}
  \caption{\textit{Top}: The \mesa\, dimensionless electron degeneracy parameter $\mathrm{\eta \sim E_F/k_BT}$, which indicates the level of degeneracy, as a function of enclosed mass for different ZAMS masses (color coded), at the tip of the RGB. \textit{Bottom}: Enclosed mass at which $\eta > 4$ (i.e. the total mass of strongly degenerate material, black dots) or $\eta > 0$ (mass of the material with some level of degeneracy, open circles) at the tip of the RGB as a function of ZAMS mass. Both panels correspond to our reference model (Z=0.02 with core overshooting during the MS).}
    \label{FIG:degeneracy}
\end{center}
\end{figure}

The level of degeneracy is determined in \mesa\, by the dimensionless electron degeneracy parameter $\mathrm{\eta \sim E_F/k_BT}$. According to \citet{paxton2011}, this parameter corresponds to the ratio of the electron chemical potential to $\mathrm{k_BT}$. In principle, any value of $\eta>0$ indicates some level of electron degeneracy, while for $\eta = 4$ the electron degeneracy pressure is roughly twice that of an ideal electron gas. In \mesa, $\eta = 4$ is used to determine whether a region of the star is degenerate or not (for example in the default TRho\_Profile plot). 
In the top panel of Fig.\,\ref{FIG:degeneracy}, we show $\eta$ as a function of enclosed mass ($m_\mathrm{r}$) for different ZAMS masses at the tip of the RGB. Only regions with $\eta>0$, i.e. close to the center, are shown. For all the models, the level of degeneracy decreases from the center outward, as expected due to the decrease in density. Even for a star with $M_{\rm ZAMS}=4.0$\,\Msun\, there is a very small level of degeneracy at the center. However, only stars with $M_{\rm ZAMS}\,\lappr\,2.0$\,\Msun\, have zones where $\eta > 4$ and are expected to ignite helium under a highly degenerate condition, i.e. experiencing a helium core flash. One can also see from this figure that stars with ZAMS masses up to $\sim1.5$\,\Msun\, have cores that converge to nearly identical levels of degeneracy. The value of $\eta$ at the center starts to slightly decrease for $M_{\rm ZAMS}=1.6$\,\Msun\, and then decreases more rapidly for larger masses, with an abrupt change between the models with $M_{\rm ZAMS}=1.8$\,\Msun\, and $1.9$\,\Msun. This is exactly the range of initial masses where we observe the abrupt decrease in the predicted sdB masses in Fig.\,\ref{FIG:range_002ov}. This is more evident in the bottom panel of Fig.\,\ref{FIG:degeneracy}, where we show $m_\mathrm{r}$ for which there is some level of degeneracy ($\eta > 0$) or strong degeneracy ($\eta > 4$), at the tip of the RGB, as a function of initial mass. It is evident from this figure that the decrease in core mass at the tip of the RGB (and therefore on the maximum sdB mass) is strongly related to the enclosed mass for which $\eta > 4$.

For low-mass progenitors ($\rm M_{\rm ZAMS}\,\lappr\,2.0-2.1$\,\Msun), the behavior for the minimum sdB masses is the same as for the maximum sdB masses, but only $\sim 0.02$\,\Msun\,below. For stars initially more massive, the level of degeneracy in the core during the RGB phase is negligible, i.e. the ignition of helium goes smoothly, and the maximum sdB mass starts to grow again. The minimum sdB mass, on the other hand, remains constant at $0.327$\,\Msun\,for initial masses up to $\sim3$\,\Msun. For larger initial masses, the helium-core mass at the TAMS is already larger than this value, and we found that helium was ignited, after removing the envelope, for any core mass we chose after the TAMS. It is highly unlikely that an sdB star can result if the envelope is removed during the MS, as there is still hydrogen in the core. Therefore, the minimum sdB mass was set to the helium-core mass at the TAMS plus the $0.01$\,\Msun\,of hydrogen envelope left, i.e. applying the `Relax Mass' process to the first model at the base of the subgiant branch. 

Even though sdBs can be produced from massive progenitors if they lose their envelope at the base of the subgiant branch, the large expansion of the radius during the RGB phase allows binary systems with a larger range of initial periods to experience mass transfer. Also, the population of sdBs in wide binaries with un-evolved companions is expected to descend from progenitors that filled their Roche lobes during the RGB phase \citep[e.g.,][]{Vos+2019,Vos20}. For the case of sdB stars that belong to close binary systems \citep[e.g.,][]{Schaffenroth22}, with orbital periods of the order of hours to a few days, the current orbital configuration can only be understood if the initial binary system experienced a common-envelope phase \citep{Paczynski76}, in which the orbital distance was dramatically reduced. It is more likely that the mass transfer process became dynamically unstable, entering a common envelope phase, if the donor filled its Roche lobe when the envelope was already deeply convective, i.e. during the RGB phase. 
Therefore, we decided to also show in Fig.\,\ref{FIG:range_002ov} the minimum sdB mass that is obtained for massive progenitors at the base of the RGB phase (dotted line). Although \mesa\,does not distinguish the different evolutionary phases, the base of the RGB can be defined assuming that a certain percentage of the envelope is already convective. Here we assumed that the star was at the base of the RGB when 1/3 of the envelope was already convective, based on the definition used in the single star evolution code (\sse) from \citet{hurley00}. The gray shaded area in Fig.\,\ref{FIG:range_002ov} represents the whole range of possible sdB masses obtained from \mesa\,for the models with $Z = 0.02$, while the dashed area highlights the region for sdBs descending from progenitors on the subgiant branch phase, which are less likely.

\begin{figure}
\begin{center}
  \includegraphics[width=0.45\textwidth]{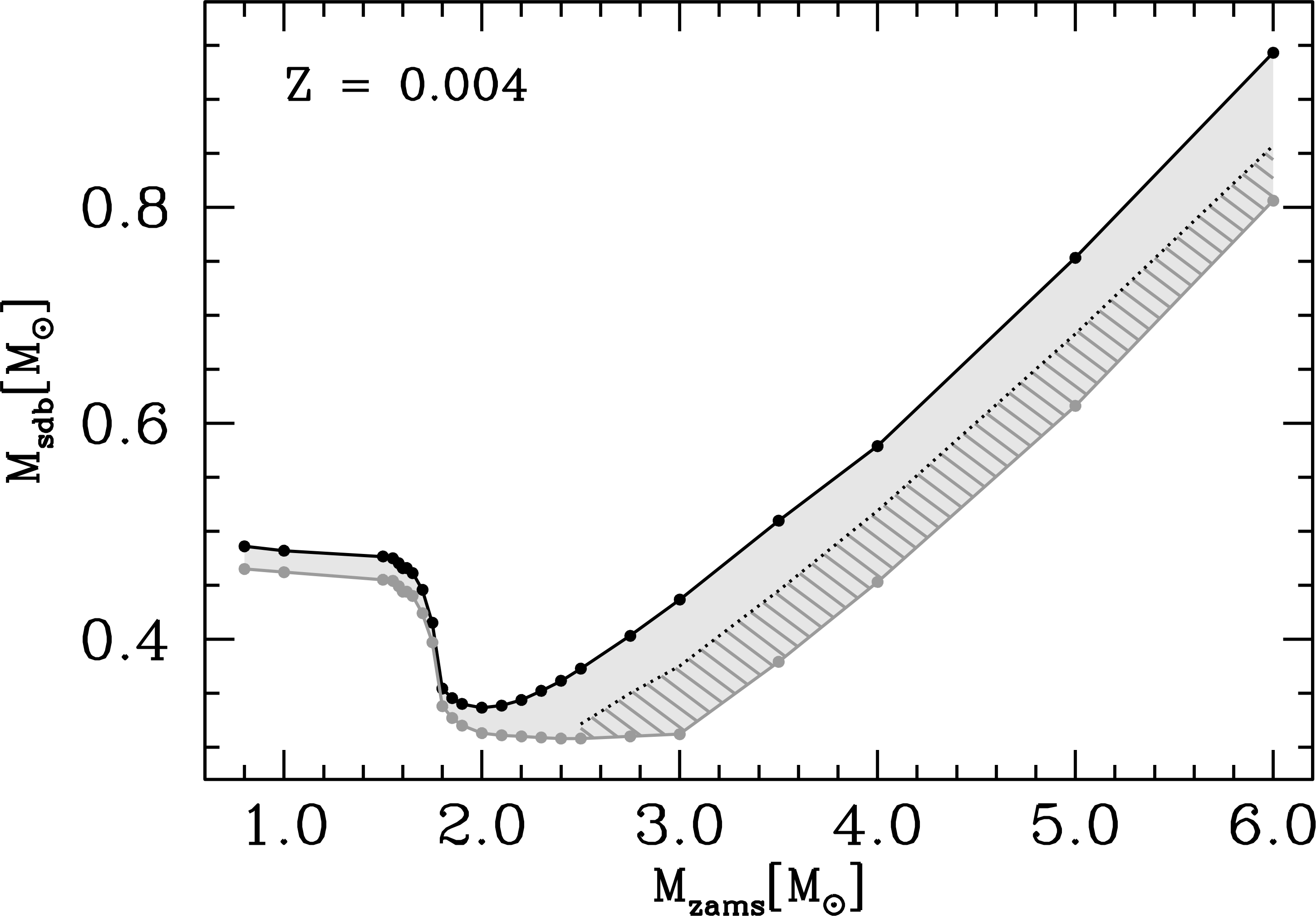}
  \caption{Same as in  Fig.\,\ref{FIG:range_002ov} but for the models with $Z = 0.004$ and core overshooting during the MS.}
    \label{FIG:range_0004ov}
\end{center}
\end{figure}

The same results are shown in Fig.\,\ref{FIG:range_0004ov} for the models with $Z = 0.004$ and overshooting, where the behaviour is very similar to the model with a larger metallicity. The tables with the results obtained from \mesa\,for our different models are presented in the appendix \ref{B:tables}.


\section{Discussion}
\label{sec:disc}

Here we analyse the effects of metallicity and overshooting. We also compare the results from \mesa\,with the ones calculated by \citet{han2002}. Finally, we show the results for the sdB lifetime as a function of sdB mass and discuss the effect this might have in the comparison between simulations and observational samples.  

\subsection{Metallicity effect}
\label{sec:met}

\begin{figure}
\begin{center}
  \includegraphics[width=0.45\textwidth]{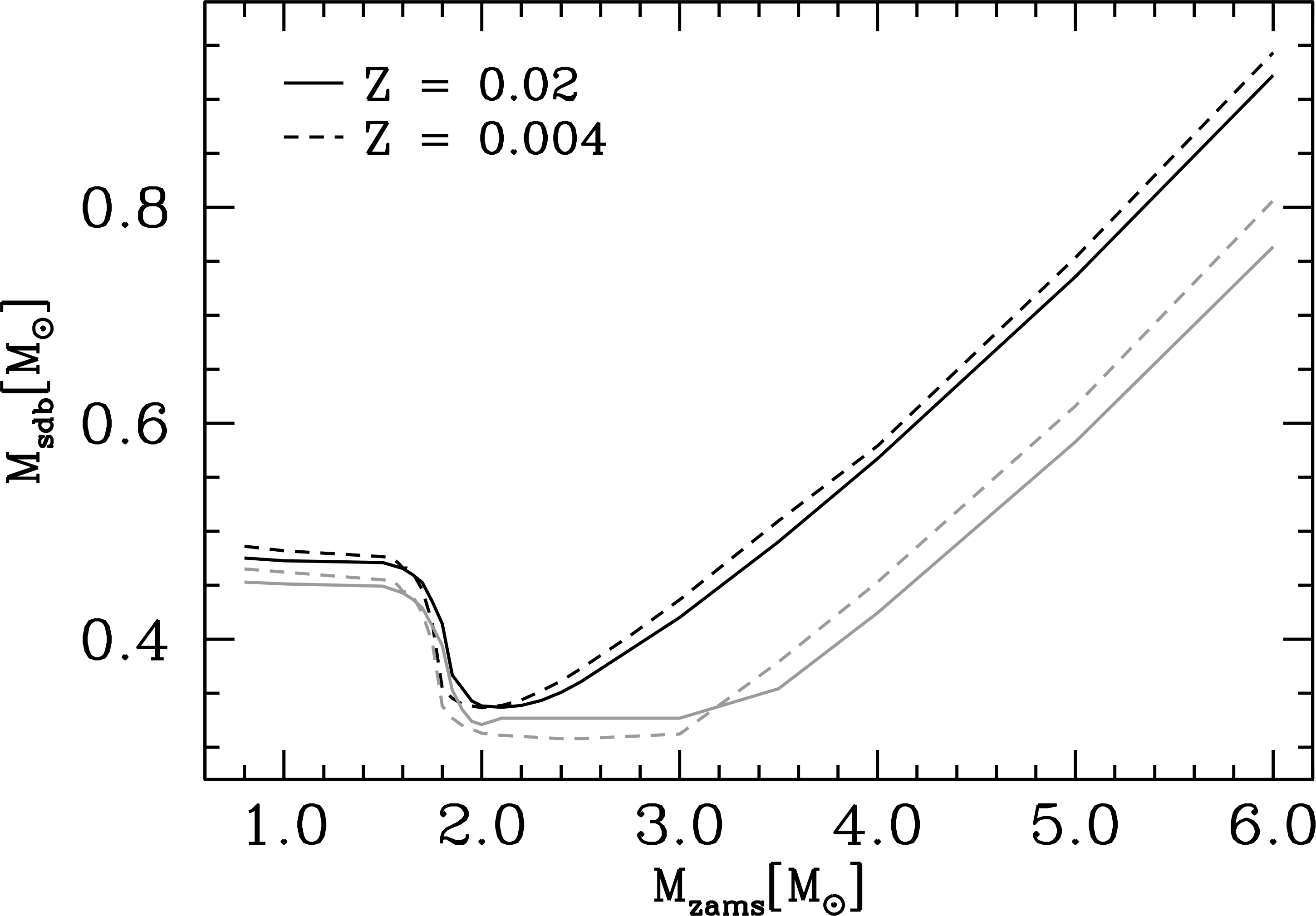}
  \caption{Maximum (black) and minimum (gray) sdB masses as a function of initial mass for the models with $Z = 0.02$ (solid lines) and with $Z = 0.004$ (dashed lines).}
    \label{FIG:zeff}
\end{center}
\end{figure}

In Fig.\,\ref{FIG:zeff} we compare the maximum (black) and minimum (gray) sdB masses as a function of the ZAMS mass for the models with $Z = 0.02$ (solid lines) and with $Z = 0.004$ (dashed lines). It is clearly seen that, for most initial masses, the whole range is slightly shifted towards larger sdB masses when a lower metallicity is assumed. This is, in part, a consequence of having more hydrogen and helium available, but is mainly related to the different opacity. Metal poor stars have a lower opacity, which allows their cores to cool easier, and therefore more mass needs to be accumulated before igniting. 
A lower metallicity implies a more massive, and therefore hotter, helium-core at the TAMS. 

For more massive progenitors ($\rm M_{\rm ZAMS}\,\gappr\,2$\,\Msun), where the level of electron degeneracy of the core during the RGB is sufficiently small to ignite helium smoothly, the maximum sdB masses are again slightly larger for the models with a lower metallicity, mainly as a consequence of the drop in the opacity, as we explained before.
However, the minimum sdB masses remain smaller for lower metallicity models in the range of $\rm M_{\rm ZAMS} \sim 2-3$\,\Msun. This is a consequence of having a hotter helium-core, where the level of degeneracy is smaller and therefore can be more compressed and heated after losing the envelope. Above $\sim\,3$\,\Msun, the minimum sdB masses are again larger for the models with lower metallicity. As it was mentioned in Sec.\,\ref{sec:sdBmasses}, any core mass we took after the TAMS ignited helium after the envelope removal for massive progenitors. Therefore, the minimum sdB masses for progenitors more massive than $\sim 3$\,\Msun\,come from the core mass at the TAMS plus the hydrogen envelope left, and lower metallicity models have a more massive core at the TAMS. 

\subsection{Comparison with the work from Han et al. 2002}
\label{sec:han}

The Eggleton code used by \citet{han2002} includes  an approach for core overshooting based on stability criteria called the `$\delta_{\rm ov}$ prescription' \citep{Pols97}, which is not available in \mesa. This makes it difficult to compare the results from both codes using models that include overshooting. Therefore, we decided to compare our results with those of \citet{han2002} only for the models without overshooting. Given the long computational time it takes to obtain the minimum sdB masses, we only calculated these for initial masses up to $2.2$\,\Msun\,in these models, following \citet{han2002}.  

\begin{figure}
\begin{center}
  \includegraphics[width=0.45\textwidth]{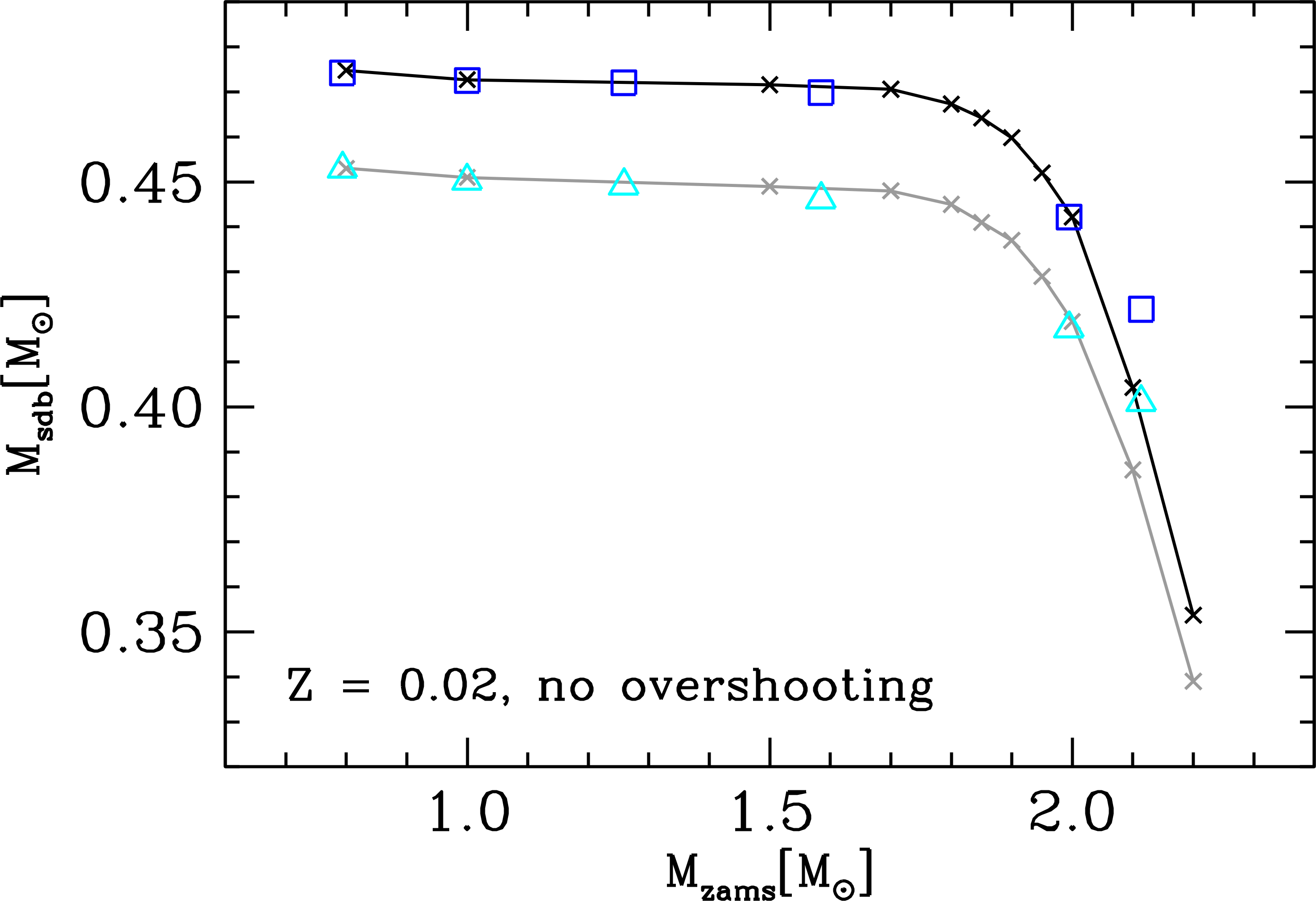}
  \caption{Maximum (black crosses) and minimum (gray crosses) sdB mass as a function of initial mass for the \mesa\,models with $Z = 0.02$ without overshooting. The values derived by \citet{han2002} for the same model are shown as blue squares and cyan triangles for the maximum and minimum sdB masses, respectively.}
    \label{FIG:han}
\end{center}
\end{figure}

In Fig.\,\ref{FIG:han} we show the comparison of the maximum (black crosses) and minimum (gray crosses) sdB masses calculated with \mesa\,for the case with $Z = 0.02$ and without overshooting, with those from \citet[][blue squares and cyan triangles, respectively]{han2002}. 
The values obtained with \mesa\,agree very precisely with those obtained by \citet{han2002} with the Eggleton code, both for the maximum and minimum sdB masses. The only exception is for the more massive progenitor calculated by \citet{han2002} for this model, i.e. the one with $\rm M_{\rm ZAMS}\sim 2.1$\,\Msun, for which we obtained smaller masses. 
However, by having a much finer grid of progenitors in the \mesa\,models, it was possible to smooth the drop in the curve that is obtained at the transition from stars that develop a degenerate core during the RGB to those that do not. This implies that using a linear interpolation within the masses derived by \citet{han2002} would have resulted in an underestimation of the sdB masses for progenitors with masses within $\sim1.6$ and $\sim2.0$\,\Msun. 

Given than \citet{han2002} did not simulate more massive progenitors in the models without overshooting, we can only conclude that the \mesa\,and Eggleton code predictions are comparably good for stars up to $\rm M_{\rm ZAMS} \sim 2.0$\,\Msun.
Despite not being able to make a direct comparison for more massive progenitors, given the difference in the overshooting prescriptions used in both codes for massive stars, we can still infer from Table\,1 in \citet{han2002} that the range of sdB masses obtained with \mesa\,is wider for stars initially more massive than $\sim2.5$\,\Msun, even if sdB descending from progenitors at the base of the RGB are considered for a more likely minimum mass.  

\subsection{The effect of core overshooting during the MS}
\label{sec:over}

\begin{figure}
\begin{center}
  \includegraphics[width=0.45\textwidth]{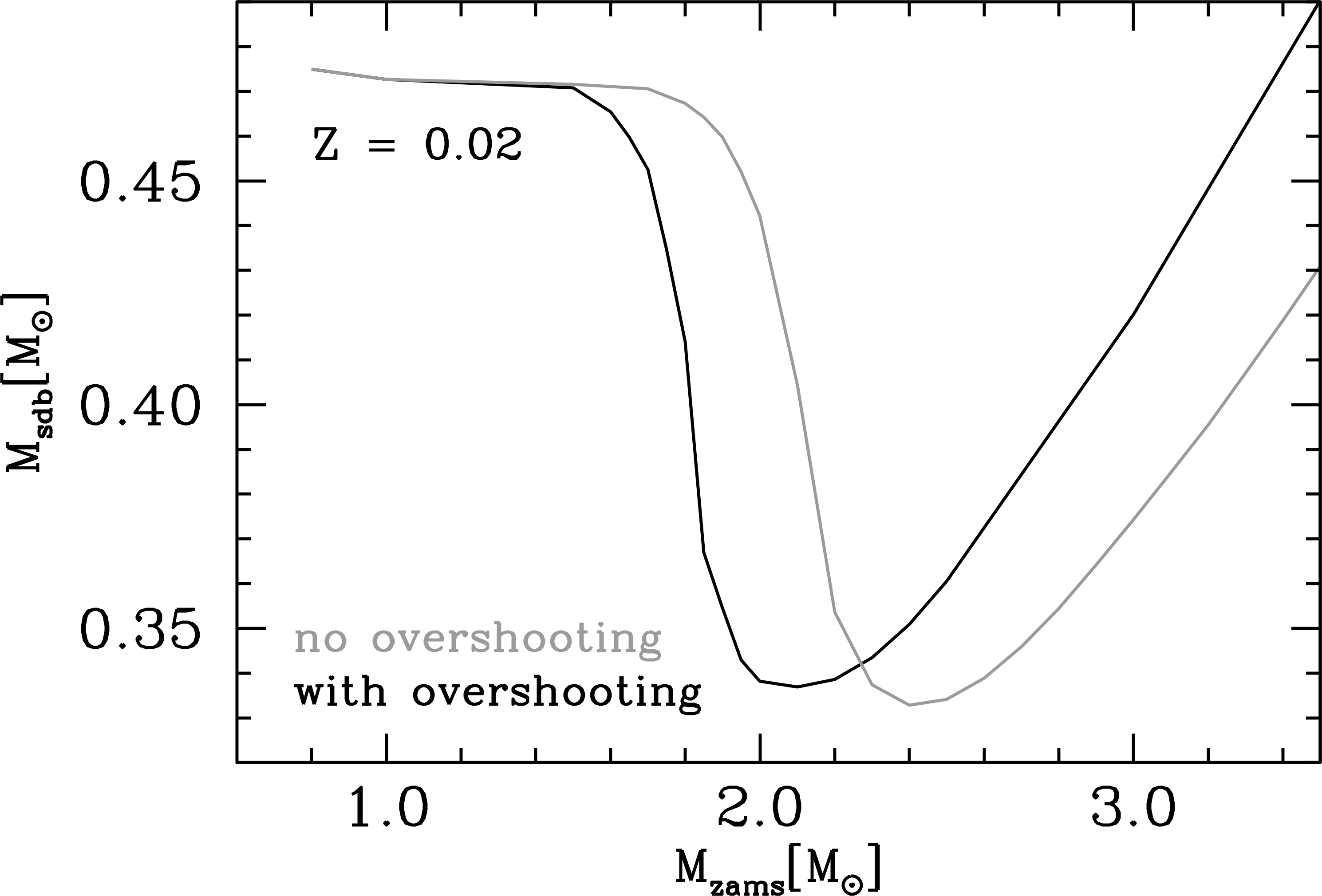}
  \caption{Maximum sdB mass as a function of initial mass for the \mesa\,models with $Z = 0.02$, with (black) and without (gray) overshooting.}
    \label{FIG:ov}
\end{center}
\end{figure}

Figure\,\ref{FIG:ov} compares the results for the maximum sdB mass as a function of initial mass for the \mesa\,models with $Z = 0.02$, with (black) and without (grey) overshooting, respectively. For stars with $\rm M_{\rm ZAMS}\,\lappr\,1.5$\,\Msun\,the maximum sdB mass is very close to the canonical value of $\sim0.47$\,\Msun, and including core overshooting during the MS does not make any difference, as expected for stars with radiative cores. For more massive stars, the shape of the two curves remains very similar, but with the curve that includes overshooting shifted to the left. 
In both cases the sdB mass decreases rapidly with increasing the initial mass, reaching a minimum value of $\sim0.33$\,\Msun\,for an initial mass of $\sim2.1$\,\Msun\,and $\sim2.4$\,\Msun\,for the cases with and without overshooting, respectively. The difference comes from the fact that considering overshooting results in a more massive and hotter core at the end of the MS, reducing the maximum initial mass for which the core becomes strongly degenerate during the RGB phase. For more massive stars, where helium is ignited in the core under non-degenerate conditions, the maximum sdB mass increases with the progenitor mass. 
A similar result was recently obtained with \mesa\,by \citet[][their Fig.\,C4]{Ostrowski2021}. 

\subsection{The duration of the sdB phase}
\label{sec:lifetime}

\begin{figure}
\begin{center}
  \includegraphics[width=0.45\textwidth]{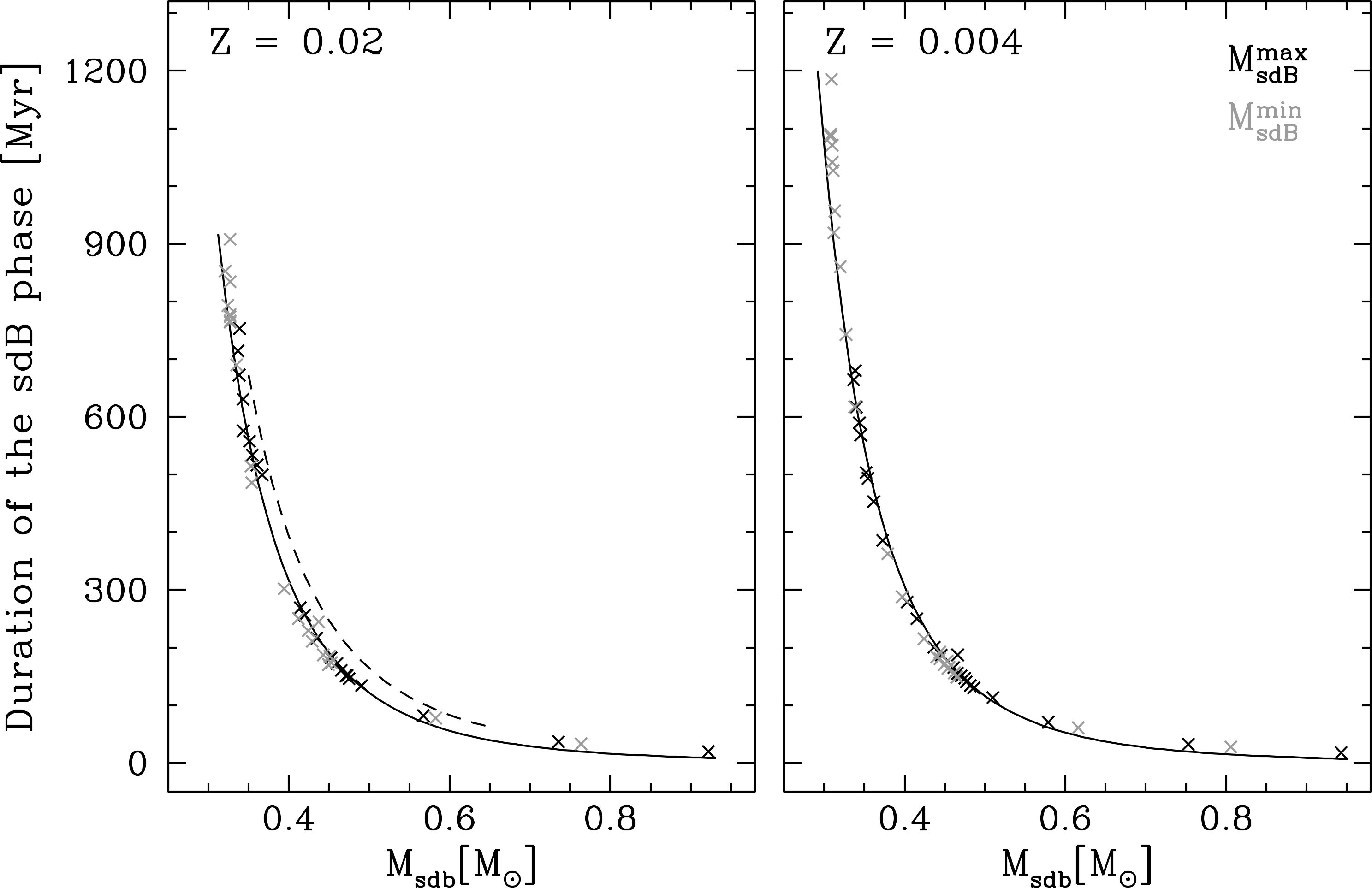}
  \caption{Duration of the sdB phase as a function of the sdB mass for the models with $Z = 0.02$ (left) and $Z = 0.004$ (right), with overshooting. The black crosses correspond to the maximum sdB masses (i.e. removing the envelope at the tip of the RGB), while the gray crosses are for the minimum sdB masses. The solid line in each panel corresponds to a linear fit to the data in a $\log-\log$ scale, while the dashed line in the left panel represents the lifetime derived by \citet{yungleson2008}.}
    \label{FIG:timesdB}
\end{center}
\end{figure}

The duration of the sdB phase is an important aspect to consider when comparing simulations with observations. In Fig.\,\ref{FIG:timesdB}, we show the duration of the sdB phase obtained from the \mesa\,models as a function of the resulting sdB mass for the models with overshooting and with a metallicity of $Z = 0.02$ (left panel) and $Z = 0.004$ (right). The black crosses correspond to sdBs descending from progenitors that lost their envelopes at the tip of the RGB, while the gray crosses represent the minimum sdB masses derived for each initial mass. Regardless of the metallicity, the duration of the sdB phase is strongly dependent on the sdB mass, decreasing for more massive sdBs, both for the maximum and minimum sdB masses. This result is not surprising, as more massive sdBs should be hotter, and therefore burn faster, which results in a shorter lifetime. 

The calculated lifetimes fit extremely well with a linear fit in the $\log \left(t_{\rm sdb} \right) - \log(M_\mathrm{sdB})$ plane (represented by the solid line in each panel) given by:
\begin{equation}
\log \left(t_{\rm sdb}[\rm Myr] \right) = -4.282 \times \log \left( \Msdb [\rm M_{\rm \odot}] \right) + 0.796,   
\end{equation}
\noindent for $Z = 0.02$ and:
\begin{equation}
\log \left(t_{\rm sdb}[\rm Myr] \right) = -4.343 \times \log \left( \Msdb [\rm M_{\rm \odot}] \right)  + 0.758,   
\end{equation}
\noindent for $Z = 0.004$. The three shortest and the three longest duration were excluded to obtain the fit for each metallicity, but they are still pretty consistent with the derived fit. The main difference between the two models is that in the low-metallicity case the least massive sdBs are $0.02$\,\Msun\,less massive than in the models with $Z = 0.02$, so that the longest lifetimes exceed $1$\,Gyr for $Z = 0.004$. However, for the same sdB mass, the low metallicity models predict a slightly shorter duration of the sdB phase. 

The dashed line in the left panel represents the sdB lifetime as a function of sdB mass derived by \citet{yungleson2008}, with the Eggleton code, for a ($0.35 - 0.65$)\,\Msun\,mass range\footnote{Although these authors did not explicitly give the assumed metallicity, they mentioned that the mass fractions of helium is $Y=0.98$ for homogeneous helium models, from which we infer $Z = 0.02$.}. While the shape of the fits is very similar, the lifetimes derived by \citet{yungleson2008} are slightly larger than the ones we obtained with \mesa. 

Even considering that low-mass sdBs spend considerably more time in the core-helium burning phase, observational evidence seems to indicate that the population is not dominated by the lower mass sdBs. From the initial mass function \citep[e.g.,][]{kroupa2001} we know that low-mass stars are much more likely to form. Therefore, most sdBs should descend from progenitors with initial masses $\lappr\,1.5$\,\Msun, resulting in an sdB mass distribution that peaks at the canonical value, with a tail towards lower mass sdB (descending from progenitors with initial masses of $\sim2-3$\,\Msun) and a smaller tail towards more massive sdBs (that descend from the more massive stars). This is consistent with the mass distribution from \citet{fontaine2012}, although it should be taken with caution due to the low number statistics (only 22 systems), but also with the most recent mass distribution derived from observational data for a larger population of sdBs in close binary systems with unevolved companions from \citet{Schaffenroth22} and for single-lined hot subdwarf stars in LAMOST from \citet{leietal2023}.  

\subsection{Implications for binary modelling}

The sdB models in this study are an important possible ingredient for the detailed modelling of the observed sdB population. Such modelling also requires performing binary population synthesis. Here, we list the key ingredients that should be considered in the binary population synthesis of sdB binaries based on our models.

1) The range of possible core masses that result in an sdB star after losing the envelope, translates into a range of radii where a star can fill its Roche lobe to become an sdB, constraining the range of initial orbital periods.

2) The lifetime of an sdB strongly depends on its mass, as we discussed above, with low-mass sdBs spending more time in this phase. 

3) The age of the population is also a crucial factor to consider in simulations that attempt to compare with observational samples of sdBs. Low-mass stars need much longer to leave the MS and become giants, i.e. possible sdB progenitors, compared to more massive stars. Also, old (low-metallicity) stars evolve faster than their younger (high-metallicity) counterparts. Therefore, a very old and low metallicity population, for example from the halo of the Milky Way, will probably have an sdB mass distribution strongly peaked at the canonical mass. This peak should become less pronounced as we move into younger populations. 

4) As we discussed in Sec.\,\ref{sec:met}, the metallicity also has a small but not negligible effect on the sdB masses. However, the metallicity strongly affects the minimum initial mass that can evolve from the MS at a given age. For example, a star of $0.8$\,\Msun\,with solar metallicity does not have enough time to leave the MS within the Hubble time, but if the same star has a very low metallicity (e.g., Z = 0.0005) it will evolve faster and will be able to leave the MS within the Hubble time (based on simulations with the \sse\, code, \citealt{hurley00}).

5) The inclusion of core overshooting during the MS, increases the resulting sdB mass for progenitors more massive than $\sim1.5$\,\Msun. However, different prescriptions for overshooting are available, without a general consensus on the more suitable. Also, the extent of the overshoot region ($f_{\rm ov}$) is another parameter that remains poorly constrained. Although $f_{\rm ov}=0.016$ seems to fit well for stars more massive than $2$\,\Msun, \citet{claret+torres16,claret+torres17,claret+torres18,claret+torres19} suggest that there is a dependence of the strength of overshooting on stellar mass, with a sharp increase between $\sim1.2$ and $\sim2$\,\Msun. This conclusion, however, is still under debate \citep[e.g.,][]{constantino+baraffe2018}.

6) The likelihood of a star to have a binary companion, which is a necessary ingredient to form sdB stars, should also be considered. There is observational evidence that the overall binary frequency is an increasing function of stellar mass \citep[e.g.,][]{Raghavan2010}.

7) It should also be taken into account whether the observed population consists of single sdBs or sdBs in wide or close binary systems. Close sdB binaries are most certainly the result of common envelope evolution, in which the envelope is rapidly ejected during the RGB phase. On the other hand, sdBs in binaries with periods of hundreds of days probably descend from a previous phase of stable mass transfer. A considerably larger timescale is needed in order to eject the envelope in the later scenario. The different timescales involved might have a measurable effect on the resulting sdB star, for example on the mass of the hydrogen envelope that is retained. For the population of single sdBs, which are expected to descend from a merger process, the internal structure of the sdBs is probably affected, and the mass distribution for this population might be completely different. 

8) SdBs in binary systems can be paired with different types of companions, that might also influence the sdB masses. For example, \citet{Schaffenroth22} suggests that the observed sdBs in close binaries show a different sdB mass distribution if they are paired with un-evolved low-mass companions, i.e. M dwarfs or brown dwarfs, or with white dwarf companions. While the sdBs with M dwarf or brown dwarf companions show a peak around the canonical mass, the peak of the distribution is shifted to lower masses for sdBs with white dwarf companions. The latter probably underwent two mass transfer phases, with the first phase caused by the evolution of the white dwarf progenitor, which must have affected the orbital period of the system before the second mass transfer phase and/or the mass of the sdB progenitor.

\vspace{0.5cm}
\noindent Having taken all these aspects of stellar population into account one might obtain a realistic mass distribution of the present Galactic sdB population.


\section{Summary}
\label{sec:sum}

In this article we have revised and refined the sdB mass range as a function of initial mass for two different metallicities, $Z = 0.02$ and $Z = 0.004$, using the stellar evolution code \mesa. We found that the lower metallicity models predict, on average, slightly more massive sdBs ($0.01-0.02$\,\Msun\,larger). The effect of including core overshooting during the MS is more evident for progenitors more massive than $\sim 1.5$\,\Msun, as expected, decreasing the maximum initial mass for which the core becomes strongly degenerate during the RGB phase, and increasing the sdB mass for progenitors that ignite helium under non-degenerate conditions. 

We also compared our results with the ranges for sdB masses derived more than two decades ago by \citet{han2002}, and found an excellent agreement for low-mass progenitors, up to $\sim2$\,\Msun, in the models without overshooting. For more massive progenitors, a direct comparison was not possible due to the different prescription for overshooting these authors used, which is not available in \mesa. However, we found that in general the \mesa\,models result in a wider mass range compared to the simulations performed by \citet{han2002} with the Eggleton code. 

The duration of the sdB phase was also calculated, finding a strong anti-correlation with the sdB mass, in agreement with previous results from \citet{yungleson2008}. The lifetime for sdBs at the lower end of the mass distribution (with $\Msdb\sim0.3$\,\Msun) is five times larger than for sdBs with the canonical mass, and at least an order of magnitude larger than for the more massive sdBs (with $\Msdb\,\gappr\,0.55$\,\Msun).

Finally, we discussed several factors that might affect the sdB mass distribution and should be considered in binary population synthesis models that aim to compare with observational samples. One of the most important factors is the age of the population where the sdBs reside, as it will constrain the minimum progenitor mass. While older populations should exhibit a strong peak at the canonical mass, corresponding to sdBs that descend from low-mass progenitors, this peak should become less pronounced or even disappear for younger populations. The evolutionary path that leads to an sdB star also needs to be considered, as we do not expect to see the same mass distribution for single (post-merger) sdBs, as for sdBs in close or wide binaries. Finally, observations seem to indicate that the sdB mass distribution is not the same for sdBs with un-evolved (MS or brown dwarf) and evolved (white dwarf) companions, at least for close binaries. Therefore, the type of companion should also be considered in population models.


\section*{Acknowledgements}
E.A., M.Z. and M.V. acknowledge support from FONDECYT (grant 1211941). 
E.A. and F.P. acknowledge partial support from Centro de Astrof\'isica de Valpara\'iso - CAV, CIDI No 17 (Universidad de Valpara\'iso, Chile) and the Postgrado en Astrofísica program (Universidad de Valparaíso, Chile).
A.B. acknowledges support for this project from the European Union’s Horizon 2020 research and innovation program under grant agreement No. 865932-ERC-SNeX.
J.V. acknowledges support from the Grant Agency of the Czech Republic (GA\v{C}R 22-34467S). 
F.P. acknowledges support from FONDECYT (grant 11180558).
This research was supported in part by the National Science Foundation under Grant No. NSF PHY-1748958.

\section*{Data Availability}
The data used in this article can be obtained upon reasonable request to the corresponding author and after agreeing to the terms of use.

\bibliographystyle{mnras}
\bibliography{paper} 

\appendix

\section{MESA inlists}
\label{A:inlists}

Here we give one example for each of the three inlist files. The star in all these samples has the same initial mass ($1$\,\Msun) and metallicity (Z = 0.004) and the envelope was extracted when the core mass was $M_\mathrm{c} = 0.452$\,\Msun. We used the \mesa\,version -r15140.

\subsection{Pre-MS to TAMS}
\label{inlist01}

\begin{center}
    \raggedright
    \ttfamily
\color{MidnightBlue}
    ! inlist to evolve a $1$\,\Msun\,star from the pre-MS to the TAMS.\\
\vspace{0.3cm}
\color{black}
    \&star\_job\ \\
\color{MidnightBlue}
    \,  ! see star/defaults/star\_job.defaults\ \\
    \,  ! begin with a pre-main sequence model\\
\color{black}
    \,    create\_pre\_main\_sequence\_model = .true.\\
\vspace{0.2cm}
\color{MidnightBlue}
    \,  ! save a model at the end of the run\\
\color{black}
    \,  save\_model\_when\_terminate = .true.\\
    \,  save\_model\_filename = \color{Green}
'sdB1.0M\_at\_TAMS.mod'\\
\vspace{0.2cm}
\color{MidnightBlue}
    \,  ! display on-screen plots\\
\color{black}
    \,  pgstar\_flag = .true.\\
\vspace{0.2cm}
/ \color{MidnightBlue} ! end of star\_job namelist\\

\color{black}
\vspace{0.3cm}
\&eos\\  
\color{MidnightBlue}
    \,  ! eos options\\
    \,  ! see eos/defaults/eos.defaults\\
\vspace{0.2cm}
\color{black} / \color{MidnightBlue} ! end of eos namelist\\

\color{black}
\vspace{0.3cm}
\&kap\\
\color{MidnightBlue}
    \,  ! kap options\\
    \,  ! see kap/defaults/kap.defaults\\
\color{black}
    \,  use\_Type2\_opacities = .true.\\
    \,  Zbase = 0.004\\
\vspace{0.2cm}
/ \color{MidnightBlue} ! end of kap namelist\\
\color{black}
\vspace{0.3cm}
\&controls\\
\color{MidnightBlue}
    \,  ! see star/defaults/controls.defaults\\
    \,  ! starting specifications\\
\color{black}
    \,  initial\_mass = 1.0 \color{MidnightBlue} ! in Msun units\\ \color{black}
    \,  initial\_z = 0.004\\
\vspace{0.2cm}
\color{MidnightBlue}
    \,  ! options for energy conservation \\
\color{black}
    \,  use\_dedt\_form\_of\_energy\_eqn = .true.\\
    \,  use\_gold\_tolerances = .true.\\
\vspace{0.2cm}
\color{MidnightBlue}
    \,  !max\_years\_for\_timestep = 5d6\\
    \,  !mesh\_delta\_coeff = 0.5\\
\vspace{0.2cm}
\color{MidnightBlue}
    \,  ! The non-default mixing parameters:\\
\color{black}
    \,  mixing\_length\_alpha = 1.8d0   \color{MidnightBlue}
! The Henyey theory of convection\\\color{black}
    \,  MLT\_option = 'Cox'\\
    \,  use\_Ledoux\_criterion = .true. \\    
\vspace{0.2cm}
\color{MidnightBlue}
    \,  ! The 'predictive mixing' scheme\\
\color{black}
    \,  predictive\_mix(1) = .true.\\
    \,  predictive\_zone\_type(1) = 'any'\\
    \,  predictive\_zone\_loc(1) = 'core'\\
    \,  predictive\_bdy\_loc(1) = 'any'\\
    \,  predictive\_superad\_thresh(1) = 0.005\\
    \,  predictive\_mix(2) = .true.\\
    \,  predictive\_zone\_type(2) = 'any'\\
    \,  predictive\_zone\_loc(2) = 'surf'\\
    \,  predictive\_bdy\_loc(2) = 'any'\\
    \,  predictive\_superad\_thresh(2) = 0.001\\
\vspace{0.2cm}
\color{MidnightBlue}
    \,  ! Core overshoot\\
\color{black}
    \,  overshoot\_scheme(1) = 'exponential'\\
    \,  overshoot\_zone\_type(1) = 'any'\\
    \,  overshoot\_zone\_loc(1) = 'core'\\
    \,  overshoot\_bdy\_loc(1) = 'top'\\
    \,  overshoot\_f(1) = 0.016\\
    \,  overshoot\_f0(1) = 0.008\\
    \,  overshoot\_mass\_full\_off(1) = 1.10\\
    \,  overshoot\_mass\_full\_on(1) = 1.30\\
\vspace{0.2cm}
\color{MidnightBlue}
    \,  !Stop at the TAMS \\
\color{black}
    \,  xa\_central\_lower\_limit\_species(1) = 'h1' \\
    \,  xa\_central\_lower\_limit(1) = 1d-4  \\
\vspace{0.2cm}
/ \color{MidnightBlue} ! end of controls namelist\\
\color{black}
\end{center}

\subsection{TAMS to the tip of the RGB}
\label{inlist02}

\begin{center}
    \raggedright
    \ttfamily
\color{MidnightBlue}
! inlist to evolve a $1$\,\Msun\,star from the TAMS to the tip of the RGB.\\
\vspace{0.3cm}
\color{black}
    \&star\_job\ \\
\color{MidnightBlue}
    \,  ! see star/defaults/star\_job.defaults\\
    \,  ! Load a previous model to run\\
\color{black}
    \,  load\_saved\_model = .true.\\
    \,  saved\_model\_name = \color{Green} 'sdB1.0M\_at\_TAMS.mod'\\
\color{MidnightBlue}
    \,  ! save a model at the end of the run\\
\color{black}
    \,  save\_model\_when\_terminate = .true.\\ 
    \,  save\_model\_filename = \color{Green} 'sdB1.0M\_at\_TRGB.mod'\\
\color{MidnightBlue}
    \,  ! Alternatively, file name for stopping at a given core mass.\\
    \,  ! save\_model\_filename ='sdB1.0M\_at\_RGB\_0.452Hecore.mod'\\
\color{MidnightBlue}
    \,  ! display on-screen plots\\
\color{black}
    \,  pgstar\_flag = .true.

/ \color{MidnightBlue} ! end of star\_job namelist\\

\vspace{0.3cm}
\color{black}
\&eos\\
\color{MidnightBlue} 
    \,  ! eos options\\
    \,  ! see eos/defaults/eos.defaults \\
\color{black} / \color{MidnightBlue} ! end of eos namelist\\

\vspace{0.3cm}
\color{black} 
\&kap\\
\color{MidnightBlue} 
    \,  ! kap options\\
    \,  ! see kap/defaults/kap.defaults\\
\color{black}
    \,  use\_Type2\_opacities = .true.\\
    \,  Zbase = 0.004d0\\ 
\color{black} / \color{MidnightBlue} ! end of kap namelist\\

\vspace{0.3cm}
\color{black}
\&controls\\
\color{MidnightBlue} 
    \,  ! see star/defaults/controls.defaults\\
    \,  ! starting specifications\\
\color{black}
    \,  initial\_mass = 1.0 \color{MidnightBlue} ! in Msun units\\ \color{black}
    \,  initial\_z = 0.004d0\\
\color{MidnightBlue} 

\vspace{0.2cm}
    \,  ! options for energy conservation \\
\color{black}
    \,  use\_dedt\_form\_of\_energy\_eqn = .true.\\
    \,  use\_gold\_tolerances = .true.\\
\color{MidnightBlue} 
    \,  !max\_years\_for\_timestep = 5d6\\
    \,  !mesh\_delta\_coeff = 0.5\\
\color{black}
    \,  photo\_interval= 100\\
\color{MidnightBlue} 
\vspace{0.2cm}
    \,  ! The non-default mixing parameters:\\
\color{black}     
    \,  mixing\_length\_alpha = 1.8d0   \color{MidnightBlue} ! The Henyey theory of convection\\\color{black}     
    \,  MLT\_option = 'Cox'\\
    \,  use\_Ledoux\_criterion = .true. \color{MidnightBlue}  \\
    
\vspace{0.2cm}
\color{MidnightBlue}
    \,  ! The 'predictive mixing' scheme\\
\color{black}     
    \,  predictive\_mix(1) = .true.\\
    \,  predictive\_zone\_type(1) = 'any'\\
    \,  predictive\_zone\_loc(1) = 'core'\\
    \,  predictive\_bdy\_loc(1) = 'any'\\
    \,  predictive\_superad\_thresh(1) = 0.005\\

    \,  predictive\_mix(2) = .true.\\
    \,  predictive\_zone\_type(2) = 'any'\\
    \,  predictive\_zone\_loc(2) = 'surf'\\
    \,  predictive\_bdy\_loc(2) = 'any'\\
    \,  predictive\_superad\_thresh(2) = 0.001\\

\vspace{0.2cm}
\color{MidnightBlue}
    \,  ! Wind in te RGB path evolution    \\
    \,  !cool\_wind\_full\_on\_T = 9.99d9\\
    \,  !hot\_wind\_full\_on\_T = 1d10\\
\color{black}
    \,  cool\_wind\_RGB\_scheme = 'Reimers'\\
\color{MidnightBlue}
    \,  !cool\_wind\_AGB\_scheme = 'Blocker'\\
    \,  !RGB\_to\_AGB\_wind\_switch = 1d-4\\
\color{black}
    \,  Reimers\_scaling\_factor = 0.25d0\\  
\color{MidnightBlue}
    \,  !Blocker\_scaling\_factor = 0.0003d0\\ 
    
\vspace{0.2cm}
\color{MidnightBlue}
    \,  !Stop at the tip of the RGB\\
\color{black}    
    \,  power\_he\_burn\_upper\_limit = 10d0\\
\vspace{0.2cm}    
\color{MidnightBlue}
    \,  ! Alternatively, stop at a specific model number (that corresponds to a fixed helium core mass)\\
    \,  ! max\_model\_number = 9850  !Model number for Mc = 0.452 \Msun.\\
\color{black}
/ \color{MidnightBlue} ! end of controls namelist\\
\color{black}
\end{center}

\subsection{Removing the envelope and evolving until the white dwarf cooling track}
\label{inlist03}

\begin{center}
    \raggedright
    \ttfamily
\color{MidnightBlue}
    ! inlist to remove the envelope for a $1$\,\Msun\,star on the RGB, at a given helium core mass (in this sample $0.452$\,\Msun), to the WD cooling track\\
\vspace{0.3cm}
\color{black}
    \&star\_job\\
\color{MidnightBlue}
    \,  ! see star/defaults/star\_job.defaults\\
    \,  ! Load a previous model to run\\
\color{black}    
    \,  load\_saved\_model = .true.\\
    \,  saved\_model\_name = \color{Green}'sdB1.0M\_at\_RGB\_0.452Hecore.mod'\\
\color{MidnightBlue}
    \,  ! save a model at the end of the run\\
\color{black}    
    \,  save\_model\_when\_terminate = .true.\\ 
    \,  save\_model\_filename = \color{Green}'sdB1.0M\_0.452Hecore\_at\_WD.mod'\\
\color{MidnightBlue}
    \,  ! display on-screen plots\\
\color{black}  
    \,  pgstar\_flag = .true.\\
\color{MidnightBlue}    
    \,  ! Relax the mass to the sdB mass\\
\color{black}     
    \,  relax\_mass = .true.\\
    \,  relax\_initial\_mass= .false.\\
    \,  new\_mass = 0.462\\
    \,  lg\_max\_abs\_mdot = -100\\  

/ \color{MidnightBlue} ! end of star\_job namelist\\

\vspace{0.3cm}
\color{black}
\&eos\\
\color{MidnightBlue}
    \,  ! eos options\\
    \,  ! see eos/defaults/eos.defaults\\

\color{Black}/ \color{MidnightBlue} \ ! end of eos namelist\\

\vspace{0.3cm}
\color{black}
\&kap\\
\color{MidnightBlue}
    \,  ! kap options\\
    \,  ! see kap/defaults/kap.defaults\\
\color{Black}    
    \,  use\_Type2\_opacities = .true.\\
    \,  Zbase = 0.004d0\\ 
/ \color{MidnightBlue} ! end of kap namelist\\

\vspace{0.3cm}
\color{black}
\&controls\\
\color{MidnightBlue}
    \,  ! see star/defaults/controls.defaults\\
\vspace{0.2cm}    
    \,  ! starting specifications\\    
\color{Black}    
    \,  initial\_mass = 1.0 \color{MidnightBlue} ! in Msun units\\
\color{Black}
    \,  initial\_z = 0.004d0\\
\color{MidnightBlue}
\vspace{0.2cm}
    \,  ! options for energy conservation \\
\color{Black}    
    \,  use\_dedt\_form\_of\_energy\_eqn = .true.\\
    \,  use\_gold\_tolerances = .true.\\
\vspace{0.2cm}
    \,  photo\_interval= 250\\
    \,  photo\_digits = 5\\
    \,  history\_interval = 1\\
\vspace{0.2cm}
 \color{MidnightBlue} \, ! Relax convergence criteria (needed during flash)\\
\color{Black}
    \,  convergence\_ignore\_equL\_residuals = .true. \\
    \,  min\_timestep\_limit = 1d-30\\
    \,  varcontrol\_target = 1d-4\\
\vspace{0.2cm}
\color{MidnightBlue}
    \,  ! The non-default mixing parameters:\\
\color{Black}     
    \,  mixing\_length\_alpha = 1.8d0  \color{MidnightBlue} ! The Henyey theory of convection\\
\color{Black}
    \,  MLT\_option = 'Cox'\\
    \,  use\_Ledoux\_criterion = .true. \\
\vspace{0.2cm}
\color{MidnightBlue} 
    \, ! The 'predictive mixing' scheme\\
\color{Black}     
    \,  predictive\_mix(1) = .true.\\
    \,  predictive\_zone\_type(1) = 'any'\\
    \,  predictive\_zone\_loc(1) = 'core'\\
    \,  predictive\_bdy\_loc(1) = 'any'\\
    \,  predictive\_superad\_thresh(1) = 0.005\\
\vspace{0.2cm}
    \,  predictive\_mix(2) = .true.\\
    \,  predictive\_zone\_type(2) = 'any'\\
    \,  predictive\_zone\_loc(2) = 'surf'\\
    \,  predictive\_bdy\_loc(2) = 'any'\\
    \,  predictive\_superad\_thresh(2) = 0.001\\
\vspace{0.2cm}
\color{MidnightBlue}     
    \,  ! Stopping criteria\\
\color{Black}     
    \,  log\_L\_lower\_limit = -3.5\\
/ \color{MidnightBlue} ! end of controls namelist\\

\end{center}

\newpage
\section{Tables}
\label{B:tables}

In the following tables we present the results obtained whit \mesa. For each initial mass ($\rm M_{\rm ZAMS}$) we listed the minimum and maximum sdB masses ($\rm M_{\rm sdB}^{\rm min}$ and $\rm M_{\rm sdB}^{\rm max}$, respectively) and the corresponding duration of the sdB phase for both limits ($\rm t_{\rm sdB}^{\rm min}$ and $\rm t_{\rm sdB}^{\rm max}$). 

\begin{table}
\centering
\caption{Model with Z = 0.02 and with overshooting. The last four values for the minimum sdB masses (highlighted with *) correspond to progenitors that lost their envelopes at the base of the subgiant branch, as explained in Section\,\ref{sec:res}.}
\begin{tabular}{cccrr}
\hline
$ \rm M_{\rm ZAMS} $ [\Msun]  & $ \rm M_{\rm sdB}^{\rm min} $ [\Msun] & $ \rm M_{\rm sdB}^{\rm max} $ [\Msun] & $ \rm t_{\rm sdB}^{\rm min} $ [Myr] & $ \rm t_{\rm sdB}^{\rm max} $ [Myr] \\
\hline \hline
0.80  & 0.453 & 0.475 & 173.8 & 146.2 \\
1.00  & 0.451 & 0.473 & 185.0 & 151.9 \\
1.50  & 0.449 & 0.471 & 170.6 & 151.4 \\
1.60  & 0.443 & 0.465 & 187.1 & 160.6 \\
1.65  & 0.437 & 0.460 & 244.8 & 172.6 \\
1.70  & 0.429 & 0.453 & 210.8 & 182.1 \\
1.75  & 0.412 & 0.435 & 250.5 & 216.0 \\
1.80  & 0.394 & 0.414 & 301.9 & 268.9 \\
1.85  & 0.352 & 0.367 & 514.4 & 499.5 \\
1.90  & 0.335 & 0.355 & 690.2 & 533.8 \\
1.95  & 0.323 & 0.343 & 793.2 & 630.3 \\
2.00  & 0.320 & 0.338 & 852.5 & 672.3 \\
2.10  & 0.326 & 0.337 & 777.2 & 714.3 \\
2.20  & 0.326 & 0.339 & 908.0 & 753.1 \\
2.30  & 0.327 & 0.343 & 773.5 & 575.8 \\
2.40  & 0.327 & 0.351 & 763.9 & 557.8 \\
2.50  & 0.327 & 0.361 & 834.4 & 516.5 \\
3.00  & 0.327 & 0.420 & 766.5 & 256.3 \\
3.50  & 0.354* & 0.490 & 485.7 & 134.0 \\
4.00  & 0.424* & 0.567 & 229.3 & 81.6 \\
5.00  & 0.583* & 0.736 & 77.9 & 36.9 \\
6.00  & 0.764* & 0.922 & 33.0 & 19.7 \\
\hline
\end{tabular}
\label{table1}
\end{table}

\begin{table}
\centering
\caption{Model with Z = 0.004 and with overshooting. As in Table\,\ref{table1}, the values with * correspond to progenitors at the base of the subgiant branch.}
\begin{tabular}{cccrr}
\hline
$ \rm M_{\rm ZAMS} $ [\Msun]  & $ \rm M_{\rm sdB}^{\rm min} $ [\Msun] & $ \rm M_{\rm sdB}^{\rm max} $ [\Msun] & $ \rm t_{\rm sdB}^{\rm min} $ [Myr] & $ \rm t_{\rm sdB}^{\rm max} $ [Myr] \\
\hline \hline
0.80 & 0.465 & 0.486 & 148.9 & 130.2 \\
1.00 & 0.462 & 0.482 & 156.0 & 133.9 \\
1.50 & 0.455 & 0.477 & 165.6 & 140.5 \\
1.55 & 0.454 & 0.475 & 163.5 & 146.7 \\
1.58 & 0.449 & 0.470 & 170.5 & 151.0 \\
1.60 & 0.444 & 0.466 & 180.8 & 155.3 \\
1.62 & 0.444 & 0.466 & 192.1 & 187.2 \\
1.65 & 0.440 & 0.461 & 183.7 & 165.3 \\
1.70 & 0.424 & 0.445 & 215.3 & 186.7 \\
1.75 & 0.397 & 0.415 & 288.0 & 249.9 \\
1.80 & 0.338 & 0.354 & 617.4 & 493.4 \\
1.85 & 0.327 & 0.345 & 742.8 & 568.6 \\
1.90 & 0.320 & 0.340 & 860.3 & 616.9 \\
2.00 & 0.313 & 0.337 & 957.0 & 664.3 \\
2.10 & 0.311 & 0.339 & 1027.1 & 679.9 \\
2.20 & 0.310 & 0.344 & 1070.9 & 589.7 \\
2.30 & 0.309 & 0.352 & 1185.2 & 503.2 \\
2.40 & 0.308 & 0.362 & 1089.8 & 453.2 \\
2.50 & 0.308 & 0.373 & 1085.7 & 385.9 \\
2.75 & 0.310 & 0.403 & 1041.2 & 278.8 \\
3.00 & 0.312* & 0.437 & 919.0 & 200.3 \\
3.50 & 0.379* & 0.510 & 362.7 & 113.2 \\
4.00 & 0.453* & 0.579 & 176.4 & 70.7 \\
5.00 & 0.616* & 0.753 & 60.9 & 32.4 \\
6.00 & 0.806* & 0.943 & 27.678 & 17.8 \\
\hline
\end{tabular}
\label{table2}
\end{table}

\begin{table} 
\centering
\caption{Models with Z = 0.02 and without overshooting. For progenitors more massive than $2.2$\,\Msun\, we did not calculate the minimum sdBs masses (see Section\,\ref{sec:over}).}
\begin{tabular}{cccrr}
\hline
$ \rm M_{\rm ZAMS} $ [\Msun]  & $ \rm M_{\rm sdB}^{\rm min} $ [\Msun] & $ \rm M_{\rm sdB}^{\rm max} $ [\Msun] & $ \rm t_{\rm sdB}^{\rm min} $ [Myr] & $ \rm t_{\rm sdB}^{\rm max} $ [Myr] \\
\hline \hline
0.80 & 0.453 & 0.475 & 173.8 & 154.0 \\
1.00 & 0.451 & 0.473 & 184.9 & 149.9\\
1.50 & 0.449 & 0.472 & 178.6 & 152.5 \\
1.70 & 0.448 & 0.471 & 181.7 & 149.4 \\
1.80 & 0.445 & 0.467 & 179.6 & 161.8 \\
1.85 & 0.441 & 0.464 & 199.4 & 162.3 \\
1.90 & 0.437 & 0.460 & 196.9 & 169.3 \\
1.95 & 0.429 & 0.452 & 213.4 & 180.6 \\
2.00 & 0.419 & 0.442 & 114.6 & 248.3 \\
2.10 & 0.386 & 0.404 & 343.1 & 295.0 \\
2.20 & 0.339 & 0.354 & 648.4 & 518.6 \\
2.30 & -      & 0.337 & -       & 732.8 \\
2.40 & -      & 0.339 & -       & 714.8 \\
2.50 & -      & 0.334 & -       & 855.2 \\
2.60 & -      & 0.339 & -       & 633.0 \\
2.70 & -      & 0.346 & -       & 583.9 \\
2.80 & -      & 0.354 & -       & 510.5 \\
2.90 & -      & 0.364 & -       & 454.2 \\
3.00 & -      & 0.374 & -       & 387.2 \\
3.10 & -      & 0.385 & -       & 349.4 \\
3.20 & -      & 0.396 & -       & 316.4 \\
3.30 & -      & 0.407 & -       & 275.1 \\
3.40 & -      & 0.419 & -       & 280.2 \\
3.50 & -      & 0.431 & -       & 222.9 \\
\hline
\end{tabular}
\label{table3}
\end{table}

\bsp	
\label{lastpage}
\end{document}